\documentclass[aps,prx,reprint,nofootinbib,superscriptaddress]{revtex4-2}
  
\usepackage{amsmath,amssymb,bm}
\usepackage{graphicx}
\usepackage{slashed}
\usepackage{xcolor}
\usepackage{hyperref}
\usepackage{booktabs}
\usepackage{orcidlink}

\newcommand{\dd}{\mathrm{d}}

\newcommand{\ket}[1]{\left|#1\right\rangle}
\newcommand{\bra}[1]{\left\langle#1\right|}
\newcommand{\braket}[2]{\left\langle #1\middle|#2\right\rangle}

\begin{document}

\title{
Theory of Particle Scattering on Superconducting Qubits: \\
From Single Devices to Quantum Arrays
}

\author{Doojin Kim\,\orcidlink{0000-0002-4186-4265}}
\email{doojin.kim@usd.edu}
\affiliation{Department of Physics, University of South Dakota, Vermillion, South Dakota 57069, USA}
\author{Kyoungchul Kong\,\orcidlink{0000-0003-4515-7303}}
\email{kckong@ku.edu}
\affiliation{Department of Physics and Astronomy, University of Kansas, Lawrence, KS 66045, USA}
\author{Myeonghun Park\,\orcidlink{0000-0002-4010-7860}}
\email{parc.seoultech@seoultech.ac.kr}
\affiliation{Institute of Convergent Fundamental Studies, Seoultech, Seoul 01811, South Korea}
\affiliation{School of Natural Sciences, Seoultech,  Seoul 01811, South Korea}

\begin{abstract}
Superconducting qubits provide microwave-scale transitions and thus offer quantum targets for direct particle scattering with extremely small energy transfer. We develop a general framework for inelastic scattering in which an incident particle exchanges an off-shell photon with a superconducting quantum circuit, inducing a transition from a lower-energy internal state of the circuit to a higher-energy one. The target dependence is encoded in a finite-momentum electromagnetic transition current and, for charge-dominated response, in a qubit transition form factor. A key feature of particle scattering is that a microwave-scale excitation does not necessarily correspond to a long-wavelength interaction, since nonrelativistic particles can transfer spatial momenta much larger than the deposited energy. The full finite-momentum device response is therefore essential. We specialize the formalism to transmon qubits, construct a realistic electrostatic response for a planar Xmon-style transmon, and apply it to photon-mediated scattering of eV-scale bosonic dark matter, including final-state Bose stimulation. We find that the resulting enhancement substantially improves the projected sensitivity at the lowest masses. Owing to the exceptionally low transition threshold of superconducting qubits, this approach opens sensitivity to dark matter in the eV-to-keV mass range, which remains largely unexplored by existing direct-detection experiments. We further show that $N$ independently operated qubits give the expected rate proportional to $N$, while suitably prepared Dicke states can in principle achieve a collective enhancement approaching $N^2$ when spatial coherence is maintained across the array. These results establish superconducting qubits as quantum targets for direct particle scattering and provide a framework for optimizing low-threshold dark-matter searches.
\end{abstract}

\maketitle

\tableofcontents
\section{Introduction}

Quantum devices have received increasing attention as sensors for particle interactions that deposit extremely small amounts of energy in matter, owing to their characteristic excitation energies that can be orders of magnitude below conventional detector thresholds~\cite{Essig:2022dfa, Cecil:2022fiv, Buchmueller:2022djy}. Superconducting qubits have already been employed as photon counters in dark-matter searches~\cite{Dixit:2020ymh}, proposed as direct sensors of coherent hidden-photon fields~\cite{Chen:2022quj}, and used to constrain dark-matter energy deposition through excess quasiparticle power~\cite{Das:2022srn, Das:2024jdz}. 
More recently, qubit-based rare-event detector concepts have exploited substrate phonons~\cite{Linehan:2025suv}, quasi-particle dynamics~\cite{Fink:2023tvb}, and parity switching as sensing channels~\cite{Ramanathan:2024hsf,Li:2025hqk}, while measurements of residual quasiparticle populations and qubit decoherence have been used to constrain dark-matter--electron scattering in transmon devices~\cite{Hochberg:2026mdl}.
Related proposals exploit quantum sensors to detect photons from radiative (rare) decays of dark matter or relic neutrinos~\cite{Dong:2025mdk}. 
These approaches demonstrate the utility of superconducting circuits as probes of weak energy deposition and electromagnetic signals at very low energies.

A complementary class of proposals considers direct particle scattering in low-threshold quantum or quantum-enabled detectors. 
Dark-matter scattering from electronic degrees of freedom in superconducting and other condensed-matter targets has been studied extensively, including formulations in terms of the many-body dielectric response~\cite{Hochberg:2021pkt, Knapen:2021run, Knapen:2021bwg, Lasenby:2021wsc, Jeong:2026buu}. 
At the device level, graphene Josephson-junction detectors have been proposed to register the minute energy deposited by dark-matter--electron scattering through a calorimetric response~\cite{Kim:2020coy}, while trapped ions can probe particle scattering through individual motional excitations or anomalous heating~\cite{Budker:2021quh}. 
More recently, coherent interactions of free electrons with quantum systems have been explored in trapped-ion quantum processors and in general quantum-optical settings, including microwave-frequency qubits~\cite{Pescoller:2026zhp, Grzesik:2026yyg}.
Related quantum-sensing concepts exploit phonons generated by particle interactions in a bulk target, including Ge-based phononic architectures proposed for low-mass dark-matter and neutrino detection~\cite{Mei:2025fxj, Mei:2025ozs}.
Such studies demonstrate that quantum technologies can provide sensitivity to extremely small energy transfers from individual particles. In these approaches, however, the quantum device generally serves to detect energy deposited into microscopic material or motional degrees of freedom, rather than treating a resolved transition between eigenstates of a superconducting qubit itself as the scattering process. 
This distinction motivates us to ask whether the qubit can instead be regarded directly as a quantum scattering target, with its internal transition forming the observable final state.

In this work, we consider particle scattering that directly induces a resolved transition between internal quantum states of the superconducting circuit. More concretely, we envision the ``2-to-2'' scattering process involving an incident (outgoing) particle $X$ ($X'$) with four-momentum $p$ ($p'$):
\begin{equation}
    X(p)+\ket a \to X'(p')+ \ket b,
    \label{eq:process}
\end{equation}
where $\ket a$ and $\ket b$ denote stationary eigenstates of a qubit system of interest and the transition is mediated by an off-shell photon. We take $\ket a$ and $\ket b$ as the ground and excited states characterized by energy eigenvalues $E_a$ and $E_b$, respectively. 
For generality purposes, $X'$ may be different from $X$, which is applicable to inelastic scattering on the particle side. 
In addition, we also formulate the scattering rate with the quantum-statistical occupation of the outgoing momentum mode explicitly retained. 
This extends the dilute-particle description to incorporate Bose stimulation or Pauli blocking for bosonic or fermionic final states, respectively, while leaving the finite-momentum electromagnetic response of the qubit unchanged.

The central theoretical object of this framework is the finite-momentum matrix element of the circuit electromagnetic current, rather than the absorption probability associated with an on-shell electromagnetic mode. The momentum transfer $q^\mu=p^\mu-p'^\mu$ carries the discrete transition energy $q^0=\omega_{q}\equiv E_b-E_a$, while its spatial component probes the geometry of the transition charge and current distribution.

This distinction is particularly relevant for particle dark matter. A microwave-scale gap allows sensitivity to energy transfers well below the meV scale, but an inelastic transition also imposes a nonzero momentum transfer. For nonrelativistic scattering, the kinematics imply that the minimum momentum transfer $Q_-$ satisfies \(Q_-\geq \omega_q/v\) for a given dark-matter speed $v$, so even a GHz-scale qubit transition can probe spatial scales substantially smaller than the overall device dimensions. This makes the finite-momentum response of the transmon an essential part of the scattering problem, rather than permitting a simple long-wavelength approximation. The observable rate is therefore controlled jointly by the level spacing and the full finite-$Q$ transition response. The formalism developed here makes this momentum matching, i.e., the overlap between the transferred momentum and the spatial Fourier components of the qubit transition charge/current distribution,  explicit through the momentum-dependent form factor or response function and applies equally to other incident particles coupled electromagnetically.

To obtain quantitative predictions for realistic devices, we further develop a complementary electrostatic-response formulation for charge-dominated interactions. Using electrostatic reciprocity, we relate the induced offset charge to an auxiliary weighting potential determined by the circuit geometry and dielectric environment. 
The resulting response kernel can be obtained from a single boundary-value problem, and its Fourier transform directly enters the scattering amplitude. This provides a practical connection between the general electromagnetic transition formalism and finite-momentum scattering calculations for realistic superconducting circuits, including the effects of conductor geometry and electrostatic screening.

Once the single-qubit response is formulated, an important question is whether an array of multiple qubits responds collectively or independently to particle scattering, particularly when the qubits are prepared in a quantum-mechanically entangled state.
In this context, we also investigate whether multiple qubits can provide a collective enhancement of the microscopic scattering rate. 
For an array initialized in the product ground state, summing over the complete set of one-excitation final states yields the expected extensive scaling with the number of qubits. By contrast, an initially excited symmetric Dicke state~\cite{Dicke:1954zz} can support superabsorption-like transitions with an $N^2$ matrix-element enhancement in the long-wavelength limit, closely related to collective absorption in engineered quantum systems~\cite{Higgins:2014cxq}. Finite momentum transfer introduces an array coherence factor that determines whether this enhancement survives for particle scattering.

As a representative application, we consider (nonthermally produced) bosonic dark matter with masses in the eV-to-keV range, a regime that remains largely inaccessible to existing direct-detection searches. 
This illustrates the interplay between the kinetic threshold and finite-momentum circuit response. 
An additional feature arises from the potentially large occupation numbers of the dark-matter momentum modes, which can lead to appreciable Bose stimulation in the final state. We find that this quantum-statistical enhancement substantially improves the projected sensitivity at low masses relative to the dilute-particle approximation employed in our earlier companion study~\cite{Kim:2026mto}, reducing the reference scattering cross section by more than 50\% for $m_\chi\lesssim 18$~eV. 
By combining a phenomenological description of the dark-matter interaction with the finite-momentum response of a realistic transmon device, we establish a concrete connection between microscopic particle scattering and superconducting-circuit detection. 
More broadly, this benchmark demonstrates how the tunable excitation thresholds and engineerable electromagnetic responses of superconducting qubits can be exploited to explore superlight dark matter in a largely uncharted mass range.

To support this logical flow, the paper is organized as follows. We begin with developing the single-qubit transition formalism in Sec.~\ref{sec:formalism}. We then briefly discuss the scaling behavior of the event rate for an array of multiple qubits and their potential collective enhancement in Sec.~\ref{sec:array}, while relegating the detailed formalism to Apps.~\ref{sec:independent_array} and \ref{sec:dicke_array}. As a concrete example, Sec.~\ref{sec:applicationsI} applies the framework to superlight dark-matter scattering, with the numerical construction of the Xmon-style electrostatic response summarized in App.~\ref{app:xmon_response}. Experimental and theoretical issues relevant to a complete sensitivity study are discussed in Sec.~\ref{sec:discussion}. Finally, Sec.~\ref{sec:conclusion} summarizes our conclusions.

\section{Setup and Formalism}
\label{sec:formalism}
\begin{figure}[t]
    \centering
    \includegraphics[width=0.48\textwidth]{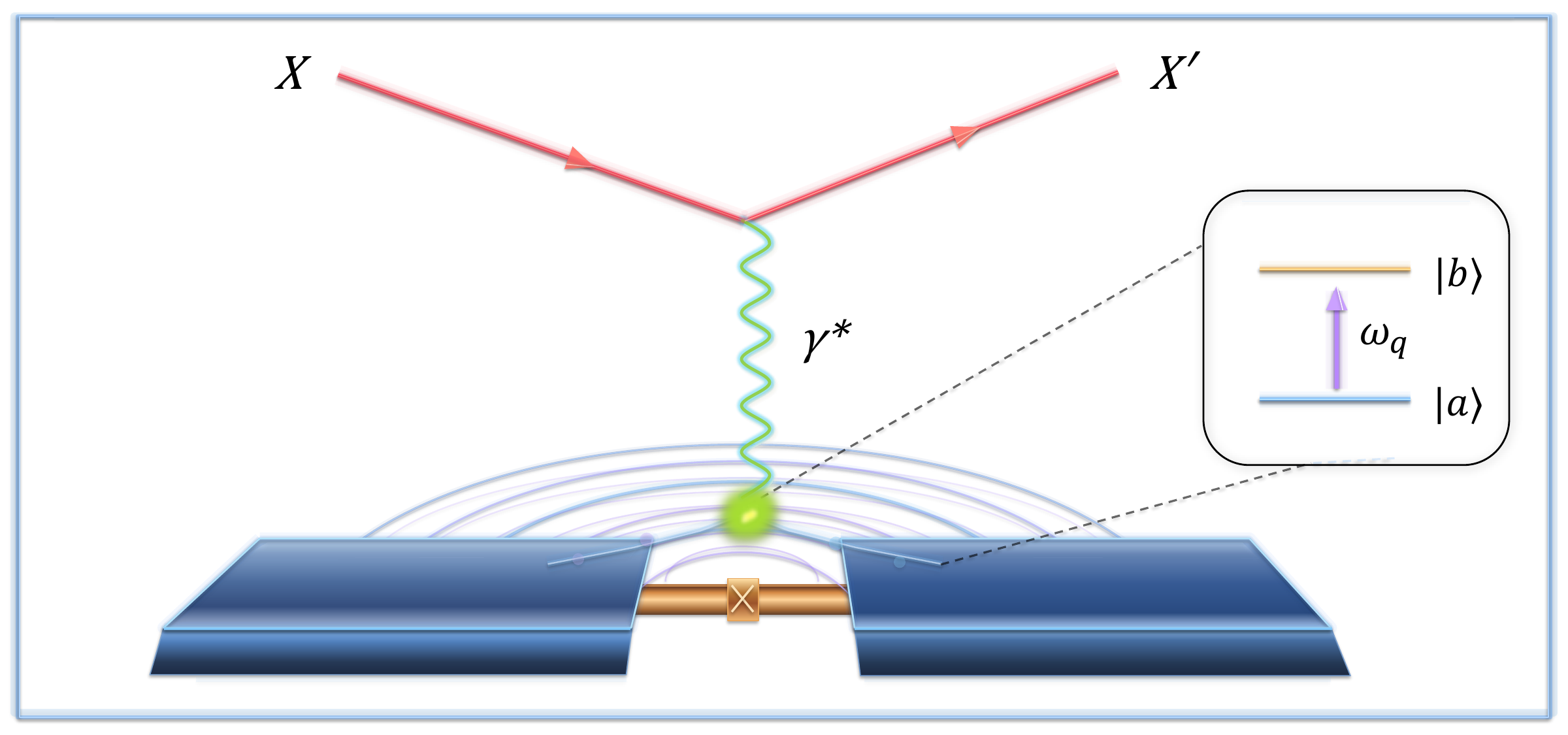}
    \caption{A schematic picture of the direct transition-scattering process $X+|a\rangle\to X'+|b\rangle$ mediated by an off-shell photon. The two electrodes schematically represent the superconducting capacitor pads, while the central orange region denotes the Josephson junction connecting them. This is for illustration only and does not represent the detailed geometry of the devices used in the numerical analysis. The momentum transfer $q^\mu=(\omega_q,\vec q)$ excites a resolved circuit transition.}
    \label{fig:schematic}
\end{figure}

We first develop the formalism to calculate the response of a given qubit system for the scattering process described in Eq.~\eqref{eq:process} (see also Fig.~\ref{fig:schematic}).
We consider a superconducting circuit with Hamiltonian $\hat H_{\rm SC}$ and stationary states
\begin{equation}
    \hat H_{\rm SC}\ket{a}=E_a\ket{a}=0,
    \qquad
    \hat H_{\rm SC}\ket{b}=E_b\ket{b},
\end{equation}
where we have the transition energy~\cite{Koch:2007hay, Krantz:2019jkw, Blais:2020wjs} 
\begin{equation}
    \omega_q= E_b-E_a>0.
\end{equation}
Here we take $E_a=0$ without loss of generality. 
The incident particle $X$ carries four-momentum $p^\mu=(E,\vec p)$ and the outgoing state $X'$ carries $p'^\mu=(E',\vec p')$. We define the momentum transfer $q^\mu$ as
\begin{equation}
    q^\mu\equiv p^\mu-p'^\mu=(\omega_{q},\vec q),
    \qquad Q\equiv |\vec q|.
    \label{eq:qdef}
\end{equation}
The center-of-mass recoil energy of a macroscopic chip is negligible throughout; the substrate and circuit support can absorb spatial momentum while the internal energy deposited in the resolved state is fixed by $\omega_{q}=E-E'$. For the momentum transfers of interest here, the corresponding wavelength remains macroscopic compared with interatomic spacings, so recoil of the rigid support is effectively recoilless and the associated Debye-Waller factor is expected to be extremely close to unity~\cite{Riedel:2012ur}.

At leading order in an electromagnetic interaction, the transition amplitude may be written as
\begin{equation}
    i\mathcal M_{ba}
    =j_X^\mu(p',p)\,[-iD_{\mu\nu}(q)]\,
    \mathcal J_{ba}^\nu(\vec q),
    \label{eq:masteramp}
\end{equation}
where $j_X^\mu$ is the incident-particle electromagnetic current and $D_{\mu\nu}$ is the photon propagator, e.g., $D_{\mu\nu}=g_{\mu\nu}/q^2$ in vacuum. In a superconducting and dielectric environment, $D_{\mu\nu}$ should be replaced by the appropriate dressed electromagnetic Green function. This substitution consistently incorporates screening, interfaces, resonant structures, and other device electrodynamics without modifying the circuit transition matrix element defined below.

To calculate the general transition cross section, we treat the qubit as a localized fixed target and ortho-normalize its stationary internal states as \(\langle a|a\rangle=\langle b|b\rangle=1\) and \(\langle a|b\rangle = 0\), while using the standard relativistic normalization for the incident and outgoing particles. For an incident particle $X$ with a speed $v=p/E$, the differential cross section for a transition $a\to b$ is
\begin{equation}
    \dd\sigma_{a\to b}
    =\frac{1}{v}\frac{1}{2E}
    \frac{\dd^3p'}{(2\pi)^3 2E'}
    (2\pi)\delta(E-E'-\omega_{q})
    \overline{|\mathcal M_{ba}|^2}.
    \label{eq:dsiggeneral}
\end{equation}
After the radial $p'$ integration,
\begin{align}
    \frac{\dd\sigma_{a\to b}}{\dd\Omega_{p'}}
    &=\frac{p'}{16\pi^2p}\,
    \overline{|\mathcal M_{ba}|^2},
    \label{eq:dsigdOmega}\\
    \frac{\dd\sigma_{a\to b}}{\dd Q^2}
    &=\frac{1}{16\pi p^2}
    \int_0^{2\pi}\frac{\dd\phi_q}{2\pi}\,
    \overline{|\mathcal M_{ba}(Q,\phi_q)|^2},
    \label{eq:dsigdQ}
\end{align}
where $\phi_q$ denotes the azimuthal orientation of $\vec q$ around the incident momentum; the average is unnecessary only when the response is azimuthally symmetric or has already been orientation averaged.
Here the kinematically allowed momentum transfer lies in
\begin{equation}
    Q_-\le Q\le Q_+,
    \qquad
    Q_-=|p-p'|,
    \quad
    Q_+=p+p'.
    \label{eq:Qrange}
\end{equation}
We emphasize that for fixed momentum transfer \(Q\), energy conservation fixes the polar angle of \(\vec q\) relative to the incident momentum, leaving only the azimuthal angle \(\phi_q\) to be integrated over. 
For an isotropic target this integral is trivial, whereas a spatially anisotropic qubit can retain a (potentially) nontrivial \(\phi_q\) dependence through its finite-momentum response. 
We will discuss this angular dependence more concretely in the context of the dark-matter application in Sec.~\ref{sec:applicationsI}.

We note that Eq.~\eqref{eq:dsiggeneral} through Eq.~\eqref{eq:dsigdQ} may be generalized to the situation where the outgoing particle $X'$ propagates in a background with non-negligible phase-space occupation. 
In this case, the transition under consideration needs to be multiplied by the quantum statistical factor,
\begin{equation}
    1+\xi_{X'}\mathcal{N}_{X'}(\vec{p}')\,, \label{eq:statfactor}
\end{equation}
where $\xi_{X'}=+1$ and $-1$ for bosons and fermions, respectively. The $\xi_{X'}=0$ choice returns the ordinary results in the classical limit (e.g., dilute particle). 
Here $\mathcal{N}_{X'}(\vec p')$ denotes the occupation number of the outgoing momentum mode, which is normalized to the corresponding number density such that
\begin{equation}
    n_{X'}=\int\frac{\dd^3 p'}{(2\pi)^3}\mathcal{N}_{X'}(\vec p').
\end{equation}
Equivalently, if the background momentum distribution $F_{X'}(\vec p')$ is normalized as $\int \dd^3p'F_{X'}(\vec p')=1$, then
\begin{equation}
\mathcal N_{X'}(\vec p')=(2\pi)^3 n_{X'}F_{X'}(\vec p'). \label{eq:NXprimfXprime}
\end{equation}
This implies that Bose stimulation (i.e., the case with $\xi_{X'}=+1$) enhances scattering into an already occupied final mode, whereas Pauli blocking (i.e., the case with $\xi_{X'}=-1$) suppresses transitions into occupied fermionic states.
The occupation of the incident mode is involved separately when the single-particle cross section is convoluted with the incident ensemble to obtain event rates. 
We will calculate the expected dark-matter event rates with this statistical factor explicitly included in Sec.~\ref{sec:applicationsI}.

\subsection{Electromagnetic transition current of a qubit} \label{sec:theoryresponse}

The electromagnetic response of the circuit is encoded in
\begin{equation}
    \mathcal J_{ba}^\mu(\vec q)
    \equiv
    \bra{b}\int \dd^3x\,
    e^{i\vec q\cdot\vec x}
    \hat J_Q^\mu(\vec x)\ket{a}
    =\bigl(\rho_{ba}(\vec q),\vec j_{ba}(\vec q)\bigr),
    \label{eq:Jtransition}
\end{equation}
where $\hat J_Q^\mu(\vec x)$ is the electromagnetic current-density operator of the qubit,\footnote{Hereafter, the subscript $Q$ on circuit and electrostatic-response quantities labels the qubit charge mode and should not be confused with the momentum-transfer magnitude $Q=|\vec q|$.} whose Fourier-transformed matrix element between $|a\rangle$ and $|b\rangle$ defines $\mathcal J_{ba}^\mu(\vec q)$.
For stationary states, current conservation implies
\begin{equation}
    q_\mu\mathcal J_{ba}^\mu=0,
    \qquad
    \omega_{q}\rho_{ba}(\vec q)
    =\vec q\cdot\vec j_{ba}(\vec q).
    \label{eq:continuity}
\end{equation}
The charge density and the longitudinal (i.e., $\vec q$-parallel) current therefore describe the same electric response, whereas the $\vec q$-transverse current contains independent magnetic and transverse-electric information.

For processes dominated by the electric/longitudinal response, a useful microscopic characterization of the circuit response is encoded in
\begin{equation}
    \rho_{ba}(\vec q)=\bra b \int \dd^3x\, e^{i\vec q \cdot \vec x}\hat\rho_{Q}(\vec x)\ket a,
    \label{eq:QTFF}
\end{equation}
where $\hat \rho_Q(\vec x)$ denotes the qubit charge-density operator, whose Fourier-transformed matrix element between $|a\rangle$ and $|b\rangle$ defines $\mathcal \rho_{ba}(\vec q)$.
For an inelastic transition, the zero-momentum charge matrix element vanishes by current conservation, i.e., \(\rho_{ba}(0)=0\) for \(b\neq a\). In the formal long-wavelength limit, the leading nonvanishing response would therefore be dipolar, but in the scattering processes of interest \(Q L\)---where \(L\) denotes the characteristic spatial extent of the qubit transition charge distribution---need not be small. We consequently retain the full finite-momentum transition form factor throughout.

For a transmon qubit, the low-energy Hamiltonian operator is given by~\cite{Koch:2007hay,Krantz:2019jkw,Blais:2020wjs}
\begin{equation}
    \hat H_{\rm SC} \to \hat H_T=4E_C(\hat n-n_g^{(0)})^2-E_J\cos\hat\varphi,
    \label{eq:transmonH}
\end{equation}
where $E_C$, $E_J$, $\hat n$, $n_g^{(0)}$, and $\hat \varphi$ are the charging energy, Josephson energy, charge number operator, effective static offset charge, and superconducting phase operator, respectively. 
Within a single collective charge mode, the spatial charge density may be parameterized as
\begin{equation}
    \hat\rho_Q(\vec x)=2e\,\hat n\,\eta_\rho(\vec x),
    \qquad
    \int \dd^3x\,\eta_\rho(\vec x)=0,
    \label{eq:rhoeta}
\end{equation}
where \(\eta_\rho(\vec x)\) describes the spatial profile of the collective transition charge associated with the transmon charge coordinate \(\hat n\). We thus identify
\begin{equation}
    \rho_{ba}(\vec q)
    =2e\,n_{ba}\,f_\rho(\vec q),
    \qquad
    n_{ba}\equiv\bra{b}\hat n\ket{a},
    \label{eq:factorize}
\end{equation}
with the geometric charge form factor
\begin{equation}
    f_\rho(\vec q)
    \equiv
    \int \dd^3x\,e^{i\vec q\cdot\vec x}\eta_\rho(\vec x).
    \label{eq:frho}
\end{equation}
The matrix element $n_{ba}$ contains the circuit selection rule and transition strength, while $f_\rho$ contains the momentum matching to the device geometry. 

In the \(E_J \gg E_C\) limit, the Hamiltonian in Eq.~\eqref{eq:transmonH} can be expanded as
\begin{equation}
\hat H_T \simeq 4E_C(\hat n-n_g^{(0)})^2-E_J
+\frac{E_J}{2}\hat\varphi^2
-\frac{E_J}{24}\hat\varphi^4,
\end{equation}
and the energy gap between the ground and excited states is given by
\begin{equation}
    \omega_q \simeq \sqrt{8E_J E_C}-E_C. \label{eq:transmonrel}
\end{equation}
At leading harmonic order, the charge operator can be written as
\begin{equation}
\hat n
\simeq
i\left(\frac{E_J}{32E_C}\right)^{1/4}
(\hat a^\dagger-\hat a)+n_g^{(0)},
\end{equation}
where $\hat a^\dagger$ and $\hat{a}$ stand for the usual ladder operators.
Consequently, for the \(0\to1\) transition, i.e., from the ground state to the first excited state,
\begin{equation}
n_{10}\equiv\langle1|\hat n|0\rangle
\simeq
i\left(\frac{E_J}{32E_C}\right)^{1/4},
~~
|n_{10}|^2
\simeq
\sqrt{\frac{E_J}{32E_C}}.  \label{eq:n10wq}
\end{equation}
For practical transmon qubits, $\omega_q$ is tuned by varying $E_J$ with $E_C$ and device geometry fixed. Therefore, it is practically convenient to express $n_{10}$ as
\begin{equation}
    |n_{10}(\omega_q)|^2 \simeq \frac{\omega_q+E_C}{16E_C}. \label{eq:n10Sqwq}
\end{equation}

To illustrate how the geometric form factor \(f_\rho(\vec q)\) can be obtained, we consider a simple toy geometry consisting of two identical thin rectangular pads lying in the \(z=0\) plane, with their centers located at \(\pm \vec d/2\). Each pad has side lengths \(L_x\) and \(L_y\), and we approximate the surface charge density as uniform over each pad. The normalized charge profile may then be written as
\begin{equation}
\eta_\rho(\vec x)
=
\frac{\delta(z)}{L_xL_y}
\left[
\Theta_+\left(x,y\right)
-
\Theta_-\left(x,y\right)
\right],
\end{equation}
where \(\Theta_\pm(x,y)\) equals unity inside the pad centered at \(\pm\vec d/2\) and vanishes outside. Using Eq.~\eqref{eq:frho}, the corresponding geometric form factor is
\begin{equation}
f_\rho(\vec q)
=
2i\sin\left(\frac{\vec q\cdot\vec d}{2}\right)
\,{\rm sinc}\left(\frac{q_xL_x}{2}\right)
\,{\rm sinc}\left(\frac{q_yL_y}{2}\right), \label{eq:recgeometry}
\end{equation}
where \({\rm sinc}(x)\equiv \sin x/x\). For two pads separated along the \(x\) direction, \(\vec d=d\,\hat x\), this reduces to
\begin{equation}
f_\rho(\vec q)
=
2i\sin\left(\frac{q_xd}{2}\right)
\,{\rm sinc}\left(\frac{q_xL_x}{2}\right)
\,{\rm sinc}\left(\frac{q_yL_y}{2}\right).
\end{equation}
The sine factor describes the relative phase between the two oppositely charged pads, while the sinc factors account for the finite spatial extent of each pad.

\subsection{Device-level electrostatic response} \label{sec:devicelevel}

For the idealized description in the previous subsection, the electromagnetic response is separated into the spatial charge profile \(f_\rho(\vec q)\) and the photon propagator. For a fabricated device, however, this separation is generally inconvenient. The electric field generated by an external particle is modified by the qubit electrodes, substrate dielectrics, ground planes, and other nearby conductors, all of which participate in the same electrostatic boundary-value problem. A more practical approach is therefore to calculate directly the offset charge induced in the transmon by an external source.\footnote{The dielectric-function and device-level electrostatic descriptions correspond to different levels of coarse-graining of the same linear electromagnetic response. The former (e.g., Refs.~\cite{Hochberg:2021pkt, Knapen:2021run, Knapen:2021bwg, Lasenby:2021wsc, Jeong:2026buu}) retains the microscopic momentum- and frequency-dependent response of a translationally invariant material, whereas the latter retains the finite device geometry and boundary conditions while representing the material response through macroscopic dielectric properties.}
Device-level electrostatic modeling, in which localized charges are coupled to a superconducting circuit by solving the electrostatic boundary-value problem for the actual conductor geometry and dielectric environment, has been widely employed in studies of charge sensitivity and charge noise in superconducting qubits (see, e.g., Refs.~\cite{Riwar:2018vpa, Christensen:2019oqh, Tennant:2021uta, Wills:2021tci}).

To formulate this device response explicitly, we first consider a classical point charge \(q_X\) located at \(\vec x\). Letting \(\mathcal{Q}_{\rm ind}(\vec x)\) denote the charge induced by this source on the transmon charge mode, we define the dimensionless response function
\begin{equation}
    \kappa_Q(\vec x)
    \equiv
    \frac{\mathcal{Q}_{\rm ind}(\vec x)}{q_X}.
    \label{eq:kappa_def}
\end{equation}
The function \(\kappa_Q(\vec x)\) is determined entirely by the device geometry and dielectric environment; for a floating two-pad transmon, \(\mathcal{Q}_{\rm ind}\) represents the differential charge associated with the two pads, whereas for an island-ground geometry it corresponds to the induced island charge. 

Directly evaluating $\kappa_Q$ would require placing the external charge at each position \(\vec x\) and repeatedly solving the electrostatic problem. 
Using the standard weighting-potential construction underlying the Shockley-Ramo theorem~\cite{Shockley:1938itm, Ramo:1939vr}, the same response can instead be obtained from a single auxiliary electrostatic problem.
For a specified device geometry, we introduce a dimensionless potential \(u_Q(\vec x)\) satisfying the generalized Laplace equation
\begin{equation}
    \vec\nabla\cdot
    \left[
        \epsilon(\vec x)\vec\nabla u_Q(\vec x)
    \right]=0,
    \label{eq:weighting_potential}
\end{equation}
where \(\epsilon(\vec x)\) is the spatially dependent permittivity. For a symmetric floating transmon, the boundary conditions may be chosen as $u_Q=\pm 1/2$ on one and the other pads, with grounded conductors fixed at \(u_Q=0\). For an island-ground transmon, one may instead take \(u_Q=1\) on the island and \(u_Q=0\) on ground. Electrostatic reciprocity then gives
\begin{equation}
    \mathcal{Q}_{\rm ind}(\vec x)
    =
    -q_X u_Q(\vec x),
\end{equation}
and hence we identify
\begin{equation}
    \kappa_Q(\vec x)=-u_Q(\vec x).
    \label{eq:kappa_uq}
\end{equation}
Thus, once the circuit geometry, dielectric constants, and conductor boundary conditions are specified, \(\kappa_Q(\vec x)\) follows from a single solution of Eq.~\eqref{eq:weighting_potential}.

We next promote the external source to a quantum particle. For a point particle with position operator \(\hat{\vec x}_X\), a small induced offset charge is given by
\begin{equation}
    \delta\hat n_g
    = \frac{\mathcal{Q}_{\rm ind}(\hat{\vec x}_X)}{2e} =
    \frac{q_X}{2e}
    \kappa_Q(\hat{\vec x}_X).
    \label{eq:dng_point}
\end{equation}
Equivalently, introducing the charge-density operator of the incident particle,
\begin{equation}
    \hat\rho_X(\vec x)
    =
    q_X\,
    \delta^{(3)}
    \left(
        \vec x-\hat{\vec x}_X
    \right), \label{eq:dng_rho_def}
\end{equation}
Eq.~\eqref{eq:dng_point} can be written as
\begin{equation}
    \delta\hat n_g
    =
    \frac{1}{2e}
    \int \dd^3x\,
    \kappa_Q(\vec x)\,
    \hat\rho_X(\vec x).
    \label{eq:dng_rho}
\end{equation}
Setting
\(
n_g^{(0)}\to n_g^{(0)}+\delta\hat n_g
\)
in the transmon Hamiltonian and expanding it to first order in the external source gives
\begin{equation}
    \hat H_T \to \hat H_T^{(0)}+\hat H_{\rm int} \approx
    \hat H_T^{(0)}
    -
    8E_C
    \left(
        \hat n-n_g^{(0)}
    \right)
    \delta\hat n_g.
\end{equation}
We therefore find the interaction Hamiltonian
\begin{equation}
    \hat H_{\rm int}
    =
    -\frac{4E_C}{e}
    \left(
        \hat n-n_g^{(0)}
    \right)
    \int \dd^3x\,
    \kappa_Q(\vec x)\,
    \hat\rho_X(\vec x),
    \label{eq:Hint_device}
\end{equation}
and the transition amplitude
\begin{equation}
    i\mathcal{M}_{ba}=-i\bra{p',b} \hat{H}_{\rm int}\ket{p,a}.
\end{equation}

We now evaluate this interaction between an initial state
\(
|p,a\rangle
\)
and a final state
\(
|p',b\rangle
\).
For \(b\neq a\),
\begin{equation}
    \langle b|
    \hat n-n_g^{(0)}
    |a\rangle
    =
    n_{ba},
\end{equation}
since the term proportional to a $c$-number \(n_g^{(0)}\) vanishes by orthogonality. The temporal component of the incident-particle electromagnetic current is defined by
\begin{equation}
    \langle p'|
    \hat\rho_X(\vec x)
    |p\rangle
    =
    e^{i\vec q\cdot\vec x}
    j_X^0(p',p),
    \qquad
    \vec q=\vec p-\vec p'.
    \label{eq:rhoX_matrix}
\end{equation}
It is therefore convenient to define
\begin{equation}
    \widetilde\kappa_Q(\vec q)
    \equiv
    \int \dd^3x\,
    e^{i\vec q\cdot\vec x}
    \kappa_Q(\vec x),
    \label{eq:kappa_fourier}
\end{equation}
and we find that the charge-response contribution becomes
\begin{equation}
    \mathcal M_{ba}
    =
    \frac{4E_C}{e}\,
    n_{ba}\,
    j_X^0(p',p)\,
    \widetilde\kappa_Q(\vec q).
    \label{eq:M_device}
\end{equation}
Thus the source dependence is contained entirely in \(j_X^0\), while the combination of the remaining factors contains the qubit transition matrix element and the complete electrostatic device response.

Combining Eq.~\eqref{eq:M_device} with the fixed-target cross section in Eq.~\eqref{eq:dsigdQ} gives
\begin{equation}
    \frac{\dd\sigma_{a\to b}}{\dd Q^2}
    =
    \frac{E_C^2|n_{ba}|^2}
         {\pi e^2p^2}
    \int_0^{2\pi}
    \frac{\dd\phi_q}{2\pi}\,
    \overline{
        |j_X^0(p',p)|^2
    }\,
    \left|
        \widetilde\kappa_Q(\vec q)
    \right|^2.
    \label{eq:dsigdq_device}
\end{equation}
This provides a direct prescription for computing the charge-induced transition rate for a specified device. If the response is averaged over the solid angle in the device or chip frame $\Omega_c$, we find
\begin{equation}
    \frac{\dd\sigma_{a\to b}}{\dd Q^2}
    =
    \frac{E_C^2|n_{ba}|^2}
         {\pi e^2p^2}
    \overline{
        |j_X^0(p',p)|^2
    }\,
    \left\langle
|\widetilde\kappa_Q(\vec q)|^2
\right\rangle_{\Omega_c},
    \label{eq:dsigdq_device_avg}
\end{equation}
where $\langle ... \rangle_{\Omega_c}$ stands for the device-orientation-averaged quantity that depends only on $Q$. For later use, we define it as
\begin{equation}
    \mathcal{K}(Q) \equiv \left\langle
|\widetilde\kappa_Q(\vec q)|^2
\right\rangle_{\Omega_c}.
\end{equation}
As briefly discussed earlier, if the outgoing particle \(X'\) has a non-negligible phase-space occupation, the integrand in Eq.~\eqref{eq:dsigdq_device} should be multiplied by the statistical factor in Eq.~\eqref{eq:statfactor}, with \(\vec p'=\vec p-\vec q\). For an isotropic final-state occupation, or after the corresponding angular average, this factor may equivalently be applied to the differential cross section in Eq.~\eqref{eq:dsigdq_device_avg}.

\section{Qubit Arrays and Rate Scaling}
\label{sec:array}

A natural way to increase the exposure of a qubit-based detector is to operate many qubits simultaneously. 
An important question is then whether the microscopic scattering rate scales simply with the number of qubits, $N$, or quantum coherence can produce a collective enhancement proportional to $N^2$.

For an array initialized in the product ground state, the answer is rather straightforward. A scattering event may excite any one of the $N$ qubits, but the corresponding one-excitation final states are mutually orthogonal. Therefore, summing over the complete set of one-excitation final states yields the ordinary linear scaling with $N$. 
This remains true even when the momentum transfer is sufficiently small that the interaction is spatially coherent across the array. Spatial coherence alone does not render the distinct one-qubit final states identical, hence does not convert the inclusive rate into an $N^2$ scaling. A detailed derivation is given in App.~\ref{sec:independent_array}.

This situation should be distinguished from quantum-enhanced searches for wave-like dark matter, such as Ref.~\cite{Chen:2023swh}, where the signal is encoded as a coherent unitary phase evolution acting simultaneously on multiple sensor qubits. As discussed in App.~\ref{sec:independent_array}, in that case, suitable entangling and readout operations can convert a collectively accumulated phase into a signal probability proportional to $N^2$. 
The process at hand is different: an individual incident particle transfers a definite energy and momentum and produces a transition into a distinct set of one-excitation final states. 
The relevant quantity is therefore the many-body transition matrix element of the scattering operator, rather than the coherent accumulation of a common phase. As a result, the $N^2$ enhancement discussed for wave-like dark matter does not apply to an array initially prepared in the product ground state considered here.

Nevertheless, a genuine enhancement beyond the extensive $N$ scaling is possible if the target qubits are prepared in an appropriate correlated many-body state. 
As discussed in our previous work~\cite{Dong:2025mdk} and in the context of superabsorption~\cite{Higgins:2014cxq}, a symmetric Dicke state with $k$ excitations can give a transition strength proportional to 
\begin{equation}
(N-k)(k+1),
\label{eq:enhance}
\end{equation} 
which approaches $N^2/4$ near the middle of the Dicke ladder, i.e., $k\to N/2$. The required Dicke state can be prepared using single-qubit gates and nearest-neighbor two-qubit gates along a chain \cite{Baertschi2019Dicke}. The enhancement in Eq.~\eqref{eq:enhance} then arises from a collective transition of the prepared state, provided the scattering phases are matched across the array. Neither state preparation nor the enhancement requires all-to-all qubit connectivity.
For the direct particle-scattering process considered here, however, realizing this limit is more restrictive than for absorption of a long-wavelength electromagnetic mode. 
The exchanged photon is off shell, and for nonrelativistic scattering the spatial momentum transfer is parametrically larger than the deposited energy, $Q\gtrsim \omega_q/v$. Hence a GHz-scale transition does not by itself imply coherence over a macroscopic array (cf. Ref.~\cite{Najera-Santos:2023jnp} for MHz-range frequency fluxonium qubits). 
The collective enhancement requires both preparation and preservation of the appropriate entangled state and spatial phase matching over the participating qubits, approximately $QD\lesssim1$, where $D$ denotes the size of the coherent region. The full many-body formalism, including the finite-momentum coherence factor, is presented in App.~\ref{sec:dicke_array}.

For the sensitivity projections in the next section, we therefore adopt the conservative and experimentally straightforward case of independently operated qubits and use the linear scaling. We briefly discuss possible quantum-enhanced operating modes based on prepared collective states in Sec.~\ref{sec:discussion}, leaving more dedicated considerations as an opportunity for future detector development.

\section{Dark-Matter Scattering}
\label{sec:applicationsI}

We now apply the formalism developed earlier to eV-scale non-thermal dark matter $\chi$, replacing $X$ in Sec.~\ref{sec:formalism} by $\chi$.
As a benchmark, we consider a complex scalar dark-matter particle carrying an effective coupling $g_\chi$,
\begin{equation}
    j_\chi^\mu
    =g_\chi\,(p+p')^\mu.
    \label{eq:dmcurrent}
\end{equation}
We then find that Eq.~\eqref{eq:dsigdq_device} becomes
\begin{eqnarray}
    \frac{\dd\sigma_{\chi,0\to 1}}{\dd Q^2}
    &=&
    \frac{E_C^2 g_\chi^2|n_{10}|^2}
         {\pi e^2 p^2}
    \int
    \frac{\dd\phi_q}{2\pi}\,
    (E+E')^2
    \left|
        \widetilde\kappa_Q(\vec q)
    \right|^2 \nonumber \\ 
    &\times& \left[ 1+\mathcal{N}_\chi (\vec p') \right],
\end{eqnarray}
where we multiply the Bose stimulation factor in Eq.~\eqref{eq:statfactor} with $\xi_\chi=+1$, since the number density of the eV-scale dark matter can be large enough to give rise to a large occupation number of the outgoing momentum mode. 
In the nonrelativistic limit,
\(E+E'\simeq2m_\chi\) and \(p\simeq m_\chi v\), yielding
\begin{equation}
    \frac{\dd\sigma_{\chi,0\to 1}}{\dd Q^2}
    =
    \frac{g_\chi^2E_C(\omega_q+E_C)}
         {4\pi e^2 v^2}
    \int
    \frac{\dd\phi_q}{2\pi}\,
    \left|
        \widetilde\kappa_Q(\vec q)
    \right|^2\left[ 1+\mathcal{N}_\chi(\vec p')\right],
    \label{eq:dsigdq_DM_device}
\end{equation}
where we have used the relation in Eq.~\eqref{eq:n10Sqwq}.

The nonrelativistic kinematics can be expressed as
\begin{equation}
    \omega_{q}=\vec q\cdot\vec v-\frac{Q^2}{2m_\chi}=Qv\cos\theta-\frac{Q^2}{2m_\chi},
    \label{eq:dmkin}
\end{equation}
where $\theta$ defines the angle between momentum transfer $\vec q$ and dark-matter velocity vector $\vec v$. The minimum speed required arises at $\cos\theta=+1$ yielding
\begin{equation}
    v_{\min}(Q,\omega_{q})
    =\frac{\omega_{q}}{Q}+\frac{Q}{2m_\chi}.
    \label{eq:vmin}
\end{equation}
For a fixed speed, the allowed momentum-transfer interval is
\begin{equation}
    Q_\pm(v,\omega_q)=m_\chi v\pm
    \sqrt{m_\chi^2v^2-2m_\chi\omega_{q}},
    \label{eq:dmQpm}
\end{equation}
and in particular
\begin{equation}
    Q_-=\frac{2m_\chi\omega_{q}}
    {m_\chi v+\sqrt{m_\chi^2v^2-2m_\chi\omega_{q}}}
    \geq\frac{\omega_{q}}{v}.
    \label{eq:Qminbound}
\end{equation}
From Eq.~\eqref{eq:dmkin}, the maximum energy transfer occurs at $\cos\theta=+1$ and $Q=m_\chi v$, yielding the necessary (and obvious) kinematic condition for the transition to occur:
\begin{equation}
    \frac12m_\chi v^2\ge \omega_{q}.
    \label{eq:dmthreshold}
\end{equation}
For the $0\to1$ transition, this gives a characteristic mass threshold
\begin{equation}
    m_\chi^{\min}\simeq
    \frac{2\omega_{q}}{v^2}.
    \label{eq:mmin}
\end{equation}
Thus the accessible dark-matter mass is controlled directly by the circuit frequency. A 5~GHz transition, $\omega_{q}\simeq20~\mu\mathrm{eV}$, has $m_\chi^{\min}\simeq 41~\mathrm{eV}$ for $v=10^{-3}$ (and about $78~\mathrm{eV}$ for $v=220~\mathrm{km/s}$), while the high-velocity halo tail extends the reach to smaller masses. The same transition satisfies $Q_-^{-1}\lesssim v/\omega_{q}\simeq9.5~\mu\mathrm m$ for $v=10^{-3}$, illustrating that the full finite charge distribution, rather than the whole-device dipole limit, is generally relevant for a conventional transmon. Extending toward $m_\chi\sim1$~eV therefore requires substantially lower-frequency circuit modes, which simultaneously lower the momentum-transfer floor but introduce thermal-occupation and noise considerations.

The event rate for a local dark-matter density \(\rho_\chi\) and laboratory-frame velocity distribution \(F_{\rm lab}(\vec v)\) is
\begin{eqnarray}
    R_\chi
    &=&
    \frac{\rho_\chi}{m_\chi}
    \int \dd^3v\,F_{\rm lab}(\vec v)\,v\, \Theta\left(\frac{1}{2}m_\chi v^2-\omega_q \right) \nonumber \\
   &\times& \int_{Q_-^2(v,\omega_q)}^{Q_+^2(v,\omega_q)} \dd Q^2\,
    \frac{\dd\sigma_{\chi,0\to1}}{\dd Q^2}, \label{eq:dmRate}
\end{eqnarray}
where the Heaviside step function explicitly enforces the kinematic threshold condition in \eqref{eq:dmthreshold}.
For a given transition, the momentum integration must use the exact interval in Eq.~\eqref{eq:dmQpm}; in particular, the inelastic lower bound in Eq.~\eqref{eq:Qminbound} should be imposed before making the heuristic estimate \(Q\sim m_\chi v\). The differential cross section entering Eq.~\eqref{eq:dmRate} is determined by the finite-momentum device response \(\widetilde\kappa_Q(\vec q)\) in Eq.~\eqref{eq:dsigdq_DM_device}, which is obtained from the electrostatic boundary-value problem described in Sec.~\ref{sec:devicelevel}. Increasing \(m_\chi\) increases the available kinetic energy, but also shifts the kinematically accessible momentum transfer toward larger \(Q\), where the response can differ substantially from its long-wavelength behavior. The useful dark-matter mass range is therefore controlled jointly by the excitation threshold and by the momentum dependence of the realistic device response.

For the rate calculation, it is convenient to combine the incident dark-matter population and the final-state Bose-stimulation factor into a dark-matter dynamic structure factor,
\begin{eqnarray}
S_\chi(\omega_q,\vec q)
&\equiv&
2\pi
\int\frac{\dd^3p}{(2\pi)^3}\,
\mathcal N_\chi(\vec p)
\left[1+\mathcal N_\chi(\vec p-\vec q)\right] \nonumber \\
&\times& \delta\!\left(
\omega_q-E_{\chi,i}
+E_{\chi, f}
\right),
\end{eqnarray}
where $E_{\chi,i} = p^2/(2m_\chi)$ and $E_{\chi,f}=(\vec p-\vec q)^2/(2m_\chi)$. As shown in Eq.~\eqref{eq:NXprimfXprime}, the occupation number of the momentum mode is related to the normalized laboratory-frame velocity distribution by
\begin{equation}
\mathcal N_\chi(\vec p)
=
(2\pi)^3\frac{\rho_\chi}{m_\chi}\frac{1}{m_\chi^3}
F_{\rm lab}\!\left(\frac{\vec p}{m_\chi}\right),
\end{equation}
where the $1/m_\chi^3$ factor arises as the Jacobian factor for $\vec v\to \vec p$.\footnote{Here we assume that for simplicity, the local dark-matter density is solely carried by $\chi$, with a negligible anti-$\chi$ abundance, as in an asymmetric dark-matter scenario.}
In terms of \(S_\chi\), the single-qubit event rate can be expressed compactly as
\begin{equation}
R_\chi
=
\frac{g_\chi^2 E_C(\omega_q+E_C)}{e^2}
\int\frac{\dd^3q}{(2\pi)^3}\,
\left|\widetilde{\kappa}_Q(\vec q)\right|^2
S_\chi(\omega_q,\vec q).
\end{equation}
The term linear in \(\mathcal N_\chi\) reproduces the usual dilute-particle result, while the quadratic term accounts for Bose stimulation into an occupied final dark-matter mode. In the limit \(\mathcal N_\chi\ll1\), the above expression is therefore reduced to the standard halo-averaged scattering rate.

For the cases where the directional information is not retained, the rate can be written as a simple factorized form. 
Defining the angle-averaged dark-matter structure factor
\begin{equation}
\overline S_\chi(\omega_q, Q)
\equiv
\frac{1}{4\pi}
\int \dd\Omega_q\,
S_\chi(\omega_q, \vec q),
\end{equation}
we find that the angle- and orientation-averaged single-qubit rate reads
\begin{equation}
R_\chi
=
\frac{g_\chi^2E_C(\omega_q+E_C)}
{2\pi^2e^2}
\int_{Q_{\min}}^{Q_{\max}} \dd Q\,Q^2\, \mathcal{K}(Q) 
\overline S_\chi(\omega_q,Q).\label{eq:nondirectionalrate}
\end{equation}
For a halo with maximum laboratory-frame speed \(v_{\max}\), the structure factor vanishes unless
\begin{equation}
v_{\min}(Q,\omega_q)
=
\frac{\omega_q}{Q}
+
\frac{Q}{2m_\chi}
\le v_{\max},
\end{equation}
so that the integral may equivalently be restricted to
\begin{eqnarray}
Q_{\min}
&=&
m_\chi v_{\max}
-
\sqrt{m_\chi^2v_{\max}^2-2m_\chi\omega_q}, \\
Q_{\max}
&=&
m_\chi v_{\max}
+
\sqrt{m_\chi^2v_{\max}^2-2m_\chi\omega_q}.
\end{eqnarray}
This one-dimensional form provides a convenient master formula for non-directional sensitivity calculations.

\subsection{Single-frequency sensitivity} \label{sec:singlefreq}

We now present the projected sensitivities at 90\% C.L. for several representative scenarios in the plane of the reference electron-scattering cross section, \(\bar{\sigma}_e\), versus the dark-matter mass, \(m_\chi\). 
Throughout this section, we use the non-directional rate in Eq.~\eqref{eq:nondirectionalrate}.
For the sensitivity estimates, we adopt the standard truncated Maxwell-Boltzmann velocity distribution~\cite{Lewin:1995rx},
\begin{equation}
F_{\rm gal}(\vec v)
=
\frac{e^{-v^2/v_0^2}}{N_{\rm esc}(\pi v_0^2)^{3/2}}
\Theta(v_{\rm esc}-v),~
F_{\rm lab}(\vec v)
=
F_{\rm gal}(\vec v+\vec v_E),
\end{equation}
where \(N_{\rm esc}\) is the normalization factor. We take $v_0=220~{\rm km/s}$, $v_E=232~{\rm km/s}$, and $v_{\rm esc}=544~{\rm km/s}$~\cite{Smith:2006ym}, and assume a local dark-matter energy density of \(\rho_\chi=0.4~{\rm GeV/cm^3}\)~\cite{Catena:2009mf, Read:2014qva}.
For each \(m_\chi\), the value of \(g_\chi\) corresponding to the projected sensitivity is translated into the conventional reference electron-scattering cross section,
\begin{equation}
\bar{\sigma}_e
=
\frac{16\pi\alpha_{\rm em}^2 g_\chi^2}{e^2q_{\rm ref}^4}
\mu_{\chi e}^2,
\end{equation}
where \(\alpha_{\rm em}\) is the electromagnetic fine-structure constant, \(\mu_{\chi e}=m_\chi m_e/(m_\chi+m_e)\) is the reduced mass of the dark-matter--electron system, and the reference momentum transfer is fixed to \(q_{\rm ref}=\alpha_{\rm em}m_e\).

We first consider an optimistic zero-background benchmark, which is intended to isolate the intrinsic sensitivity associated with direct dark-matter scattering on the qubit from detector-specific backgrounds. For a vanishing expected background, we define the 90\% C.L. sensitivity by requiring \(N_{\rm sig}=2.3\) signal events. We consider exposures of
\begin{equation}
    \mathcal{E}=1,\;10^3,\;10^6~{\rm qubit}\cdot{\rm yr}, \label{eq:refexpose}
\end{equation}
where \(\mathcal{E}=N T\) denotes the product of the number of active qubits and the live time. 
These benchmark exposures are chosen to span qualitatively different detector scales rather than as arbitrary logarithmic intervals. The \(1~{\rm qubit\cdot yr}\) benchmark illustrates the reach attainable even with a single device operated over an extended period. 
The $10^3~{\rm qubit\cdot yr}$ benchmark represents an intermediate-scale array, as superconducting quantum systems containing $10^3-10^4$ physical qubits have already been realized~\cite{Castelvecchi:2023ibm,King:2022dqa}.
The $10^6~{\rm qubit\cdot yr}$ benchmark instead represents a long-term prospect; million-qubit-scale superconducting systems are contemplated both in industrial roadmaps~\cite{Rini:2024qec,Chow:2021qia} and in architectural studies~\cite{Saadatmand:2024erv} of fault-tolerant superconducting quantum computers. These efforts illustrate that such device counts, while technologically demanding, lie within the scale contemplated for future quantum hardware. 
We emphasize that the exposure does not necessarily involve a number of physical qubits operated simultaneously; comparable exposure may also be accumulated through smaller modular arrays and longer live times.

For simplicity, we assume the ordinary $N$ scaling as mentioned in Sec.~\ref{sec:array}.
Throughout this analysis, the qubit transition frequency is fixed to \(f_{q}=5~{\rm GHz}\), corresponding to
\(
    \omega_{q}=2\pi f_{q}\simeq 20.7~\mu{\rm eV}.
\)
We emphasize that here $T$ should be interpreted as the ``fiducial'' live time during which the qubits are sensitive to particle scattering. In practice, the measurement may consist of repeated initialization, interrogation, and readout cycles. In the rare-event limit relevant for the sensitivity projections, the expected number of scattering events depends only on the accumulated live time and is (nearly) insensitive to how this time is partitioned among individual interrogation windows, provided that a transition occurring during a live window is detected with approximately unit efficiency. 
Finite excited-state relaxation, readout efficiency, and initialization/reset dead time can be incorporated into an effective live time or, equivalently, an overall detection efficiency.\footnote{This search scheme should be distinguished from coherent-field searches, for which the transition amplitude can build coherently during an interrogation interval and the signal probability therefore depends explicitly on the coherence time squared.} 

As the first benchmark for the device response, we use the simple two-pad geometry introduced in Sec.~\ref{sec:formalism}. Each pad is taken to be a thin rectangle of size
\(
    L_x=L_y=200~\mu{\rm m},
\)
with a gap of \(100~\mu{\rm m}\), corresponding to a center-to-center separation \(d=300~\mu{\rm m}\). In the uniform surface-charge approximation, the corresponding charge form factor is given by Eq.~\eqref{eq:recgeometry}. For vacuum photon propagation, the device-response function introduced in Sec.~\ref{sec:formalism} is then obtained analytically as
\begin{equation}
    \widetilde\kappa_Q^{\rm vac}(\vec q)
    =
    -\frac{e^2}{2E_C}\,
    \frac{f_\rho(\vec q)}{Q^2}.
    \label{eq:kappa-toy}
\end{equation}
We take \(E_C/(2\pi)=250~{\rm MHz}\). For \(f_{q}=5~{\rm GHz}\), the transmon relation in Eq.~\eqref{eq:transmonrel} then gives \(E_J/E_C\simeq55.1\), and hence \(|n_{10}|^2\simeq 1.31\).
In evaluating the rate, we retain the full finite-momentum dependence of Eq.~\eqref{eq:kappa-toy} and average \(|\widetilde\kappa_Q(\vec q)|^2\) over the direction of \(\vec q\).

To assess the impact of a more realistic electromagnetic environment, we also construct the response for an Xmon-style planar transmon, taking the geometry of Ref.~\cite{Barends:2013zvp} as a representative design. 
We adopt a cross-shaped island with strip width and coplanar gap \(S=W=24~\mu{\rm m}\) and arm length \(L=160~\mu{\rm m}\). Unlike the analytic two-pad approximation, this geometry explicitly includes the surrounding ground plane and the resulting nonuniform charge accumulation near conductor edges. 
The corresponding electrostatic response \(\widetilde{\kappa}_Q(\vec q)\) is obtained numerically following the procedure detailed in App.~\ref{app:xmon_response}. For this benchmark, we obtain \(C_\Sigma\simeq70~{\rm fF}\), corresponding to \(E_C/(2\pi)\simeq275~{\rm MHz}\); at \(f_q=5~{\rm GHz}\), this gives \(E_J/E_C\simeq46\) and \(|n_{10}|^2\simeq1.2\). The numerical response therefore captures the finite conductor geometry and screening by the coplanar ground plane, while not attempting to model the complete chip environment, such as control lines, readout resonators, or the package.

\begin{figure}[t]
    \centering
    \includegraphics[width=0.48\textwidth]{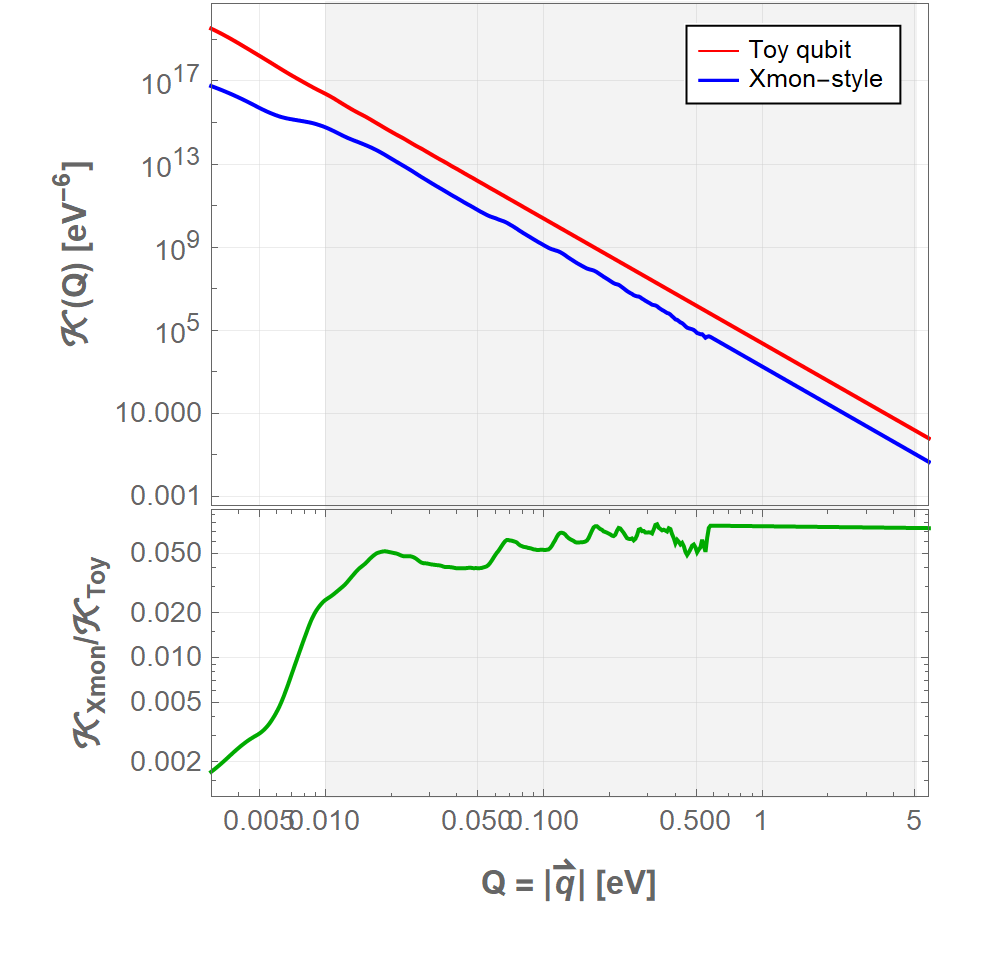}
    \caption{Comparison of the orientation-averaged squared electrostatic response $\mathcal{K}(Q)$ for the analytic two-pad toy model and the numerical Xmon-style benchmark. The lower panel shows the ratio $\mathcal{K}_{\rm Xmon}/\mathcal{K}_{\rm Toy}$. The realistic Xmon response is suppressed relative to the unscreened toy model, particularly at small momentum transfer, reflecting the effects of the surrounding coplanar ground plane and the nonuniform induced-charge distribution. The gray-shaded region indicates the range of $Q$ values relevant to our sensitivity projections.}
    \label{fig:response_comparison}
\end{figure}

To provide some intuition for the impact of the realistic device environment, Fig.~\ref{fig:response_comparison} compares the orientation-averaged response of the numerical Xmon benchmark with that of the analytic two-pad model. The Xmon response is systematically smaller over the momentum range relevant to our analysis, with the strongest suppression occurring at small \(Q\). This behavior is qualitatively expected: the analytic two-pad model describes an unscreened dipole-like charge configuration in vacuum, whereas the Xmon geometry includes a nearby coplanar ground plane and the associated redistribution of induced charge near conductor edges. At larger momentum transfer, where the scattering begins to resolve more local features of the conductor geometry, the ratio increases to the few-percent level and exhibits mild structure associated with the finite spatial profile of the device. We stress that this comparison should not be interpreted as a pure screening factor, since the two benchmarks differ in geometry and circuit parameters; rather, it illustrates the size of the correction that can arise when replacing a simple analytic form factor by a more realistic finite-device electrostatic response.

\begin{figure}[t]
    \centering
    \includegraphics[width=0.48\textwidth]{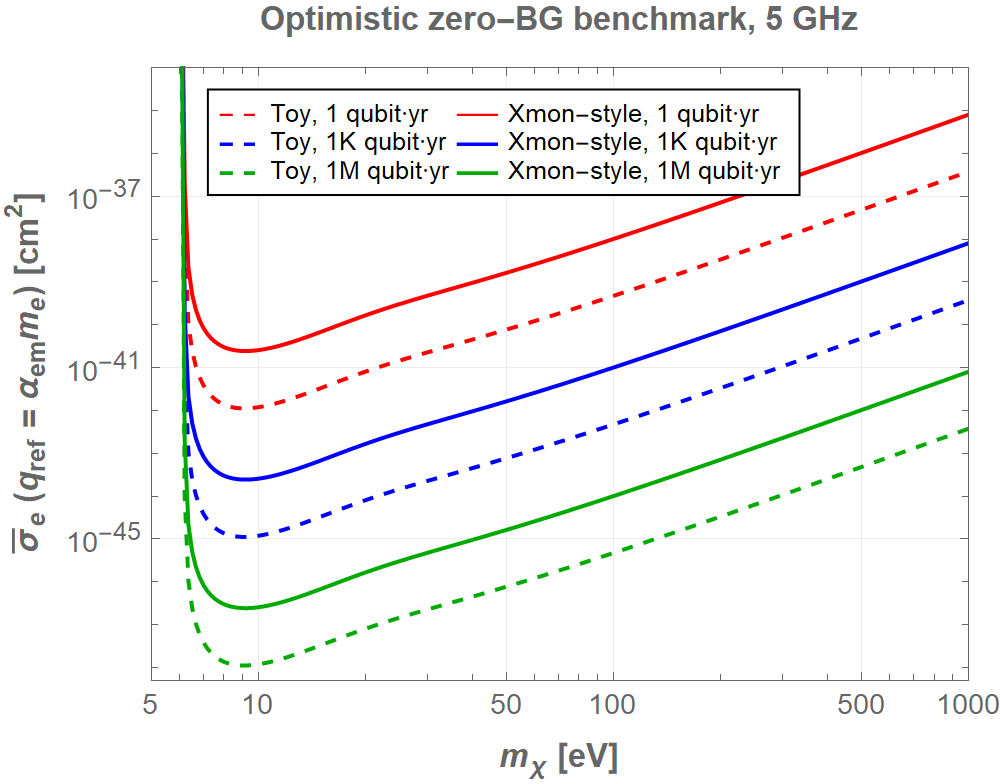} 
    \includegraphics[width=0.48\textwidth]{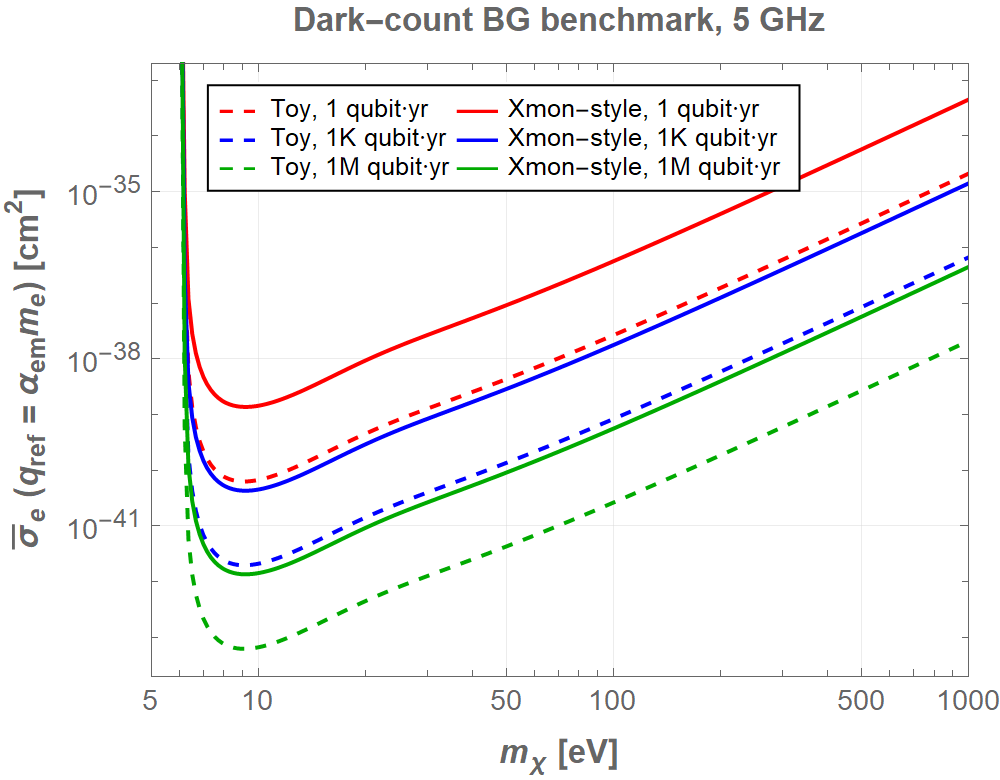}
    \caption{Projected 90\% C.L. sensitivities to the reference dark-matter--electron scattering cross section \(\bar{\sigma}_e\), defined at \(q_{\rm ref}=\alpha_{\rm em}m_e\), for a transmon transition frequency \(f_{q}=5~{\rm GHz}\). The upper panel shows the optimistic zero-background benchmark, for which the sensitivity is defined by \(N_{\rm sig}=2.3\). The lower panel includes an effective residual dark-count rate of \(b_{\rm dark}=10^4\) events/(qubit$\cdot$yr), with the corresponding 90\% C.L. signal upper limit evaluated assuming a known Poisson background. Red, blue, and green curves correspond respectively to exposures of \(1\), \(10^3\), and \(10^6~{\rm qubit}\cdot{\rm yr}\). Dashed curves use the analytic two-pad toy geometry with vacuum photon propagation, while solid curves employ the numerical Xmon-style electrostatic response including the finite conductor geometry and screening by the surrounding coplanar ground plane. See the text for more details.}
    \label{fig:dmapp}
\end{figure}

The resulting projected sensitivities are shown in the upper panel of Fig.~\ref{fig:dmapp}. For each exposure, we show both the analytic two-pad vacuum benchmark (dashed lines) and the numerical Xmon response (solid lines). The former illustrates the reach inferred from the simplified geometric form factor and vacuum propagator, whereas the latter provides a more realistic estimate of the finite-momentum electrostatic response of a planar transmon. In both cases, increasing the exposure from \(1\) (red lines) to \(10^3\) (blue lines) and \(10^6~{\rm qubit}\cdot{\rm yr}\) (green lines) improves the zero-background sensitivity linearly with exposure, while the low-mass boundary remains fixed by the kinematic requirement for exciting the \(5~{\rm GHz}\) transition.

We emphasize that the sensitivity projections in the upper panel of Fig.~\ref{fig:dmapp} also explicitly show the impact of final-state Bose stimulation especially at low dark-matter masses. Our earlier companion study~\cite{Kim:2026mto} employed the dilute-particle approximation, yielding conservative sensitivity estimates in this regime. 
The Bose-stimulation factor enhances the scattering rate and thus improves the sensitivity projections. 
More quantitatively, the $\bar\sigma_e$ is reduced by more than 50\% for \(m_\chi\lesssim 18\) eV, reaching an improvement of more than an order of magnitude at \(m_\chi=10\) eV. At higher masses, the phase-space occupation decreases and the Bose-stimulation effect becomes negligible, and the resulting projected limits approach their dilute-particle counterparts. 

In a realistic implementation, qubit excitations unrelated to the signal can generate false-positive events. Relevant sources include residual thermal population of the excited state, imperfect state preparation and readout, nonequilibrium quasiparticles, and environmental ionizing radiation. Residual excited-state populations in transmons are known to correspond to effective temperatures above the refrigerator base temperature, indicating imperfect thermalization even in dilution refrigerators~\cite{Jin:2015xlk}. Ionizing radiation can additionally produce quasiparticle and charge-burst events in superconducting-qubit substrates~\cite{Martinis:2020fxb,Bratrud:2025}. On the other hand, raw single-shot readout errors need not translate directly into detector backgrounds. Repeated quantum-nondemolition measurements have been demonstrated to provide exponential rejection of false-positive events with only a linear increase in measurement time~\cite{Dixit:2020ymh}. We therefore interpret the experimentally relevant background as the residual dark-count rate after initialization, repeated readout and event validation, rather than the raw error probability of an individual qubit measurement.

To set a representative scale for this residual background, we take guidance from recent measurements of superconducting qubits operated in a low-radiation underground environment. Reference~\cite{Bratrud:2025} reports a minimum single-qubit charge-jump rate of approximately \(0.19~{\rm mHz}\) with surrounding lead shielding, corresponding to roughly \(6\times10^3\) charge-burst events per qubit$\cdot$yr. These charge jumps should not be identified one-to-one with false excitations in the present search: some may not induce a detectable qubit transition and spatially correlated events may be rejected, while other sources such as nonequilibrium excitations and residual readout errors can contribute additional backgrounds. Nevertheless, the measured rate provides a useful empirical scale for environmental disturbances in superconducting-qubit devices. We therefore adopt
\(
b_{\rm dark}=10^4~{\rm events/(qubit}\cdot{\rm yr)}
\)
as a representative effective residual dark-count benchmark and take
\begin{equation}
B=b_{\rm dark}\mathcal E,
\end{equation}
where the exposures \(\mathcal{E}\) is given in Eq.~\eqref{eq:refexpose}.
For comparison, the thermal contribution alone for a \(5~{\rm GHz}\) qubit operated at an effective temperature of \(10~{\rm mK}\), with an interrogation time of \(100~\mu{\rm s}\), is only \({\cal O}(10)\) events per qubit$\cdot$yr. Thus, our benchmark is intended to capture residual instrumental and environmental backgrounds beyond the irreducible thermal population, while remaining anchored to experimentally observed disturbance rates in superconducting-qubit systems.

The lower panel of Fig.~\ref{fig:dmapp} shows the projected sensitivities including the benchmark residual dark-count rate introduced above. Compared with the optimistic zero-background case, the reach is weakened substantially, particularly at large exposures, because the expected background grows linearly with \(\mathcal E\). 
Nevertheless, increasing the exposure from \(1\) to \(10^3\) and \(10^6~{\rm qubit}\cdot{\rm yr}\) continues to provide a substantial gain in sensitivity. As in the upper panel, the dashed curves correspond to the two-pad vacuum benchmark, while the solid curves use the electrostatic response of the Xmon-style transmon; the latter therefore illustrate the reach after accounting simultaneously for a more realistic finite-momentum device response and residual detector backgrounds.

\medskip

It is useful to place the projected reach in the context of existing direct searches for dark-matter--electron scattering. At present, terrestrial constraints on the generic reference cross section \(\bar{\sigma}_e\) do not extend into the eV-to-keV mass range considered here.
Among the lowest-threshold searches, the QROCODILE experiment using superconducting nanowire single-photon detectors has probed dark-matter masses down to approximately \(30~{\rm keV}\)~\cite{QROCODILE:2024nqm}, while the GLIMPSE experiment using graphene Josephson junction device technology would explore the $m_\chi$ region down to the sub-keV scale~\cite{Kim:2020coy}.\footnote{Existing precision-force measurements have also been reinterpreted to constrain light dark-matter--nucleon scattering~\cite{Matsumoto:2025rcz,Liu:2025jlx}.}
Searches employing astrophysical boosting, such as solar-reflected dark matter, have recently extended xenon-based sensitivity to masses as low as \(4.6~{\rm keV}\)~\cite{XENON:2025dpl}.
Thus, when \(\bar{\sigma}_e\) is regarded simply as a phenomenological parameter describing photon-mediated dark-matter--electron scattering, the parameter space explored by the qubit detector lies below the mass reach of current terrestrial searches.

More generally, astrophysical and cosmological considerations can constrain the viable parameter space for light dark matter, depending on its properties and production history~~\cite{Ali-Haimoud:2015pwa,Chang:2019xva,Lehmann:2020lcv,Alvey:2020xsk,Ali-Haimoud:2021lka,Nguyen:2021cnb}.
A particularly restrictive case arises if the dark matter is identified with a minimal millicharged particle carrying electric charge \(\epsilon e\). In this case, stellar-cooling constraints from red-giant and horizontal-branch stars exclude much of the parameter space corresponding to the sensitivities shown here~\cite{Davidson:2000hf,Fung:2023euv}. The red-giant constraint has recently been revisited using detailed stellar-evolution calculations and remains extremely stringent for sub-\(100~{\rm keV}\) millicharged particles~\cite{Fung:2023euv}. These bounds, however, assume that the same light state and electromagnetic interaction relevant in terrestrial experiments remain operative under stellar conditions. Concrete extensions have been proposed in which environmental effects, such as density-dependent effective masses or couplings, suppress particle production in stellar interiors and thereby evade the standard cooling bounds~\cite{DeRocco:2020xdt}. Direct detection-based terrestrial searches therefore provide a complementary probe that is less dependent on assumptions about the behavior of the new sector in stellar environments.

\subsection{Frequency-multiplexed array} \label{sec:multiplex}

An attractive feature of qubit-based detection is that different qubits in an array need not be operated at the same transition frequency. We therefore consider a frequency-multiplexed configuration and compare it with the single-frequency benchmark discussed above. In both cases, the total exposure is fixed to \(1000~{\rm qubit}\cdot{\rm yr}\), and we use the electrostatic response of the Xmon-style qubit. To isolate the effects of the frequency-multiplexed search strategy, we assume the optimistic zero-background limit. The single-frequency configuration consists of \(1000\) qubits operated at \(5~{\rm GHz}\) for one year, whereas the multiplexed configuration contains \(125\) qubits at each of \(f_{q}=3,4,\ldots,10~{\rm GHz}\), again operated for one year.\footnote{The 3--10~GHz range considered here should be understood as a theoretical benchmark exploiting the tunability and design flexibility of transmon qubits. The original Xmon devices of Ref.~\cite{Barends:2013zvp} were tested over approximately 4--6~GHz, while later developments covered the 3--7~GHz range~\cite{Verjauw:2022qxp, Sah:2024xmon}.} Denoting by \(R_\chi^{(f)}\) the single-qubit event rate at transition frequency \(f\)~GHz, the expected signal yields in the two configurations are
\begin{eqnarray}
    N_{\rm sig}^{(5)}
    &=&
    (1000~{\rm qubit}\cdot{\rm yr})\,R_\chi^{(5)},\\
    N_{\rm sig}^{\rm multi}
    &=&
    (125~{\rm qubit}\cdot{\rm yr})
    \sum_{f=3}^{10} R_\chi^{(f)} .
\end{eqnarray}
For a given \(m_\chi\), the rates from the individual frequency channels are therefore combined before imposing the 90\% C.L. zero-background criterion \(N_{\rm sig}=2.3\).

\begin{figure}
    \centering
    \includegraphics[width=0.48\textwidth]{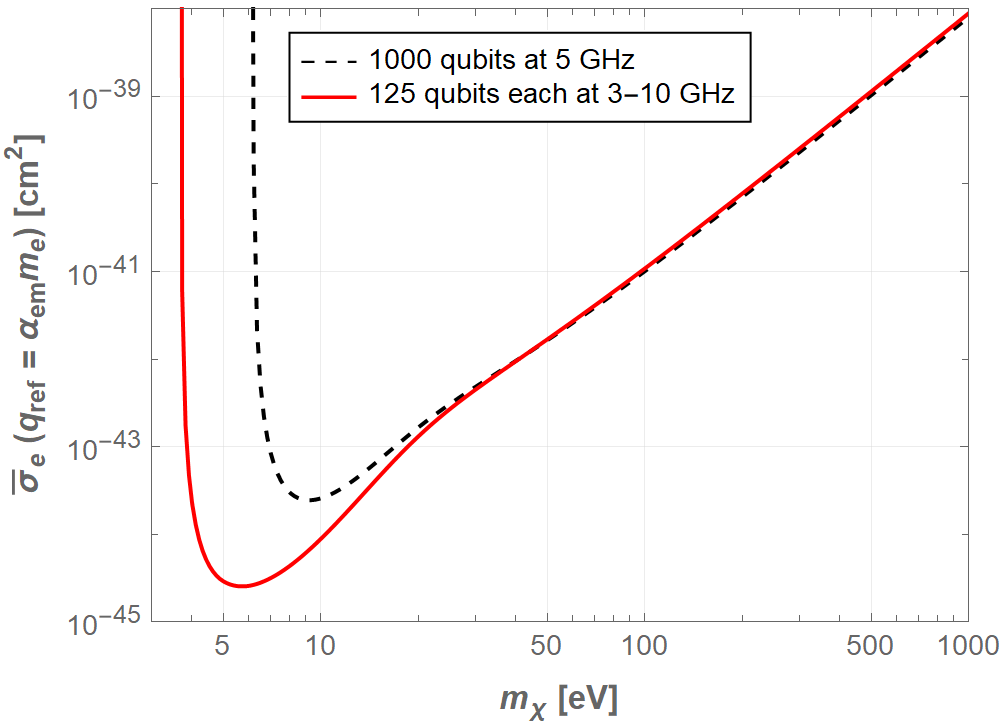}
    \caption{Projected 90\% C.L. sensitivity for two array configurations with the Xmon-style qubits in the optimistic zero-background limit, each with a total exposure of \(1000~{\rm qubit}\cdot{\rm yr}\). The black dashed curve corresponds to \(1000\) qubits operated at \(5~{\rm GHz}\) for one year, while the red solid curve corresponds to a frequency-multiplexed array with \(125\) qubits at each of \(3,4,\ldots,10~{\rm GHz}\), also operated for one year. Signal yields from all frequency channels are combined before imposing the \(N_{\rm sig}=2.3\) criterion. The multiplexed configuration extends the low-mass reach to \(m_\chi\simeq3.7~{\rm eV}\) and provides stronger sensitivity below the crossover at \(m_\chi\simeq42.6~{\rm eV}\).}
    \label{fig:DMmultiplex}
\end{figure}

The resulting sensitivities are shown in Fig.~\ref{fig:DMmultiplex}. The multiplexed array extends the kinematic reach down to \(m_\chi\simeq3.7~{\rm eV}\), set by the \(3~{\rm GHz}\) channel, and provides stronger sensitivity than the \(5~{\rm GHz}\)-only configuration up to \(m_\chi\simeq42~{\rm eV}\). Notably, the multiplexed configuration also performs better around the mass range where the \(5~{\rm GHz}\) detector itself has its best sensitivity. For example, at \(m_\chi=13~{\rm eV}\), the summed rate from the eight frequency channels is approximately \(14.6\) times the single-qubit \(5~{\rm GHz}\) rate. Since the multiplexed array contains eight times fewer qubits in each frequency channel, it outperforms the single-frequency configuration whenever
\begin{equation}
    \sum_{f=3}^{10}R_\chi^{(f)}
    >
    8\,R_\chi^{(5)}.
\end{equation}
At \(m_\chi=13~{\rm eV}\), this condition is satisfied: the \(3\), \(4\), and \(5~{\rm GHz}\) channels contribute approximately \(69\%\), \(22\%\), and \(6.9\%\) of the total multiplexed signal, respectively. The lower-frequency channels benefit from a smaller excitation threshold and hence a larger accessible halo phase space, while also probing smaller momentum transfers where the Xmon-like response is larger. As \(m_\chi\) increases, this kinematic advantage becomes less important, and the two configurations become comparable when \(\sum_f R_\chi^{(f)}\simeq8R_\chi^{(5)}\), which occurs numerically at \(m_\chi\simeq42.6~{\rm eV}\). Above this crossover, concentrating the full array at \(5~{\rm GHz}\) becomes modestly more favorable. This comparison illustrates that frequency multiplexing can substantially broaden the accessible mass range without necessarily sacrificing the peak sensitivity.

\section{Discussion and Outlook}
\label{sec:discussion}

The framework developed above identifies several directions for extending the reach of direct particle scattering on superconducting qubits.
A distinctive feature of this detection mode is that the circuit transition energy and the spatial momentum transferred by the incident particle are parametrically different quantities.
For nonrelativistic dark matter, the kinematic relation
$Q_-\geq \omega_{q}/v$ shown in Eq.~\eqref{eq:Qminbound}
implies that even a microwave-scale excitation can resolve spatial structure well below the overall dimensions of a conventional transmon.
Rather than constituting a limitation of the approach, this makes the finite-momentum electromagnetic response an intrinsic detector observable.
The transition form factor introduced in Sec.~\ref{sec:theoryresponse}, or equivalently the device-level response function developed in Sec.~\ref{sec:devicelevel}, provides a systematic way to describe this momentum matching without relying on a whole-device dipole approximation.

Our numerical Xmon-style qubit benchmark already incorporates an important subset of the electrodynamic effects that arise in a fabricated device, including finite conductor geometry, nonuniform charge accumulation, and screening by the surrounding coplanar ground plane.
A natural next step is to extend this treatment to the complete chip environment.
Within the current formulation this can be accomplished either by solving for the full electromagnetic Green function,
including dielectric interfaces, resonators, control structures, package boundaries, and superconducting screening, or by generalizing the reciprocity construction to the relevant device modes.
The present electrostatic calculation captures the dominant longitudinal charge response, whereas a complete treatment of transverse electric and magnetic currents would further enlarge the class of particle interactions that can be studied.
Importantly, such refinements modify the detector response without changing the underlying scattering formalism.

The array structure provides two complementary routes toward larger signal yields.
The most direct strategy is simply to increase the number of independently operated qubits, for which the inclusive scattering rate scales extensively as $N$.
This scaling does not require spatial coherence across the array and therefore remains applicable even when the transferred momentum resolves the separation between different qubits.
In addition, the frequency of different qubits need not be identical.
The frequency-multiplexed example of Sec.~\ref{sec:multiplex} demonstrates that distributing a fixed total exposure among several transition frequencies can broaden the accessible dark-matter mass range and improve the sensitivity over a substantial low-mass interval.
This suggests that the qubit-frequency distribution itself may be treated as a detector-design parameter and optimized jointly with the electromagnetic response and background environment.

A qualitatively different possibility is provided by quantum-engineered many-body target states.
As shown in Sec.~\ref{sec:array} and detailed in Apps.~\ref{sec:independent_array} and \ref{sec:dicke_array}, spatial coherence alone does not convert an initially unexcited product state into an $N^2$ target; the mutually orthogonal one-qubit final states retain the ordinary inclusive $N$ scaling.
For a suitably prepared Dicke state, however, the many-body transition matrix element can itself be enhanced,
reaching $(N-k)(k+1)\sim N^2/4$ near the middle of the Dicke ladder.
For finite momentum transfer, this enhancement is controlled by the array coherence factor $C_N(\vec q)$ defined in Eq.~\eqref{eq:array_coherence} and therefore requires both quantum-state preparation and spatial phase matching over the coherently participating region.
The resulting $N^2$ regime should thus be viewed as a genuine quantum-enhanced operating mode beyond the baseline independent-qubit detector, with the required state preparation, frequency uniformity, and device compactness providing concrete targets for future circuit architectures.

\begin{figure}[t]
    \centering
    \includegraphics[width=0.48\textwidth]{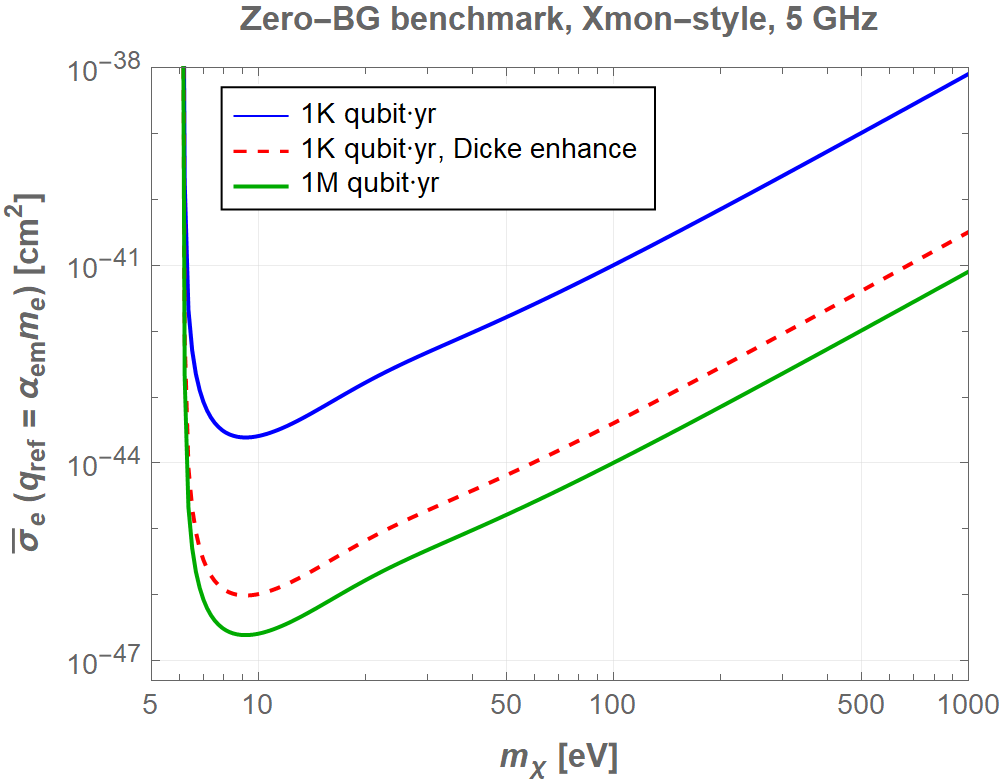}
    \caption{Projected zero-background sensitivity to the reference electron cross section \(\bar{\sigma}_e\) for an Xmon-style \(5~{\rm GHz}\) transmon benchmark. The solid blue curve shows the baseline sensitivity for \(10^3\) qubit\(\cdot\)yr of independent-qubit exposure, while the dashed red curve illustrates the same exposure assuming an ideally prepared Dicke state with collective enhancement and perfect spatial coherence. For comparison, the solid green curve shows the sensitivity of \(10^6\) qubit\(\cdot\)yr of independent qubits.}
    \label{fig:dicke_sensitivity}
\end{figure}

To illustrate the potential impact of such collective enhancement, Fig.~\ref{fig:dicke_sensitivity} compares the projected sensitivity of independent and collectively prepared qubit arrays for the same benchmark setup considered above. We take an Xmon-style planar transmon with a transition frequency of 5~GHz, assume vanishing background, and consider $10^3$ qubit$\cdot$yr of exposure. 
For independent qubits, the total rate scales linearly with the number of participating devices. For an ideally prepared Dicke state in the maximally enhanced configuration, by contrast, the collective transition factor becomes $N^2/4+N/2$, with the finite-momentum coherence factor taken to be \(C_N(\vec q)=1\) for illustration, as described in Sec.~\ref{sec:array}. For $N=10^3$, this corresponds to an effective rate enhancement of approximately $2.5\times10^2$ relative to $10^3$ independent qubits (solid blue line). 
Consequently, the Dicke-enhanced $10^3$ qubit$\cdot$yr sensitivity (dashed red line) approaches that of an independent-qubit detector with $10^6$ qubit$\cdot$yr exposure (solid green line), differing only by an $\mathcal O(1)$ factor in the ideal coherence limit. 
The comparison is not intended as a near-term experimental projection, but rather to illustrate the substantial reduction in physical detector exposure that collective quantum preparation could in principle provide.

Anisotropic target responses can encode directional information about the Galactic dark-matter wind and produce characteristic daily modulation~\cite{Boyd:2022tcn}; directional detection has also been proposed in superconductors through the angular distribution of quasi-particle excitations~\cite{Hochberg:2021ymx} and through the angular correlation between the incident dark-matter and the two-dimensional detector plane~\cite{Kim:2020coy, Jeong:2026buu}. 
In the present case, the directionality instead originates from the finite-momentum response of the qubit device itself.
For a generic circuit geometry, $\widetilde{\kappa}_Q(\vec q)$ depends on the orientation of $\vec q$ as well as its magnitude.
The convolution of this anisotropic response with the Galactic dark-matter velocity distribution can therefore retain information about the orientation of the detector relative to the dark-matter wind.
A fixed device in the laboratory may consequently exhibit a daily modulation as the Earth rotates.
Unlike the total-rate projections presented in Sec.~\ref{sec:applicationsI}, exploiting this information would require retaining the full angular dependence of the device response rather than orientation averaging it.
Engineered arrays with deliberately anisotropic or phase-patterned geometries could further enhance such directional sensitivity.
A quantitative optimization of this possibility is left for future work.

The transmon should also not be regarded as an exact two-level target.
The electromagnetic transition-current formulation applies directly to higher excited states and to other circuit modes by replacing the corresponding matrix element
$\mathcal{J}_{ba}^{\mu}(\vec q)$.
For a weakly anharmonic transmon the $0\to1$ transition provides the simplest and typically dominant benchmark for a linear charge coupling, while higher transitions, quasiparticle excitations, phonons, and other noncomputational modes can become relevant as the deposited energy increases.
The same language therefore provides a route toward a more general response tensor for superconducting quantum devices.

The dark-matter example considered here should accordingly be viewed as one concrete realization of a broader scattering framework.
Our sensitivity results are expressed in terms of the phenomenological reference cross section $\bar{\sigma}_e$, while additional constraints associated with specific ultraviolet realizations can be imposed separately, as discussed in Sec.~\ref{sec:singlefreq}.
More generally, different incident particles and microscopic electromagnetic currents can be incorporated by replacing $j_X^\mu$, e.g., exothermic relic neutrino or dark-matter scattering, while different circuit architectures enter through the corresponding finite-momentum transition response.
The separation between particle physics, device electrodynamics, and quantum-state engineering is therefore useful not only for interpreting the present benchmarks, but also for systematically optimizing future superconducting-qubit scattering experiments.

\section{Conclusions}
\label{sec:conclusion}

We have developed a general framework for direct particle scattering into resolved excitations of superconducting qubits.
The central object is the finite-momentum electromagnetic transition current $\mathcal{J}_{ba}^{\mu}(\vec q)$, which cleanly separates the microscopic interaction of the incident particle from the internal response of the quantum circuit.
For charge-dominated transitions, we introduced a qubit transition form factor and showed that, because an inelastic transition carries no net monopole charge, its zero-momentum charge matrix element vanishes.
The relevant detector response is therefore intrinsically momentum dependent.
In addition, the formulation accommodates quantum-statistical effects in the outgoing particle states, including Bose stimulation and Pauli blocking, independently of the electromagnetic transition response of the qubit.

For transmon qubits, we derived the corresponding fixed-target scattering cross section and described the electromagnetic response in two complementary ways.
An analytic geometric form factor provides a transparent description of an idealized two-pad transmon, while a device-level electrostatic response directly incorporates the finite geometry and electrostatic boundary conditions of a realistic circuit.
Applying the latter construction to an Xmon-style geometry allowed us to include finite conductor dimensions, nonuniform induced charge, and screening by the surrounding ground plane in the scattering calculation.

A central kinematic result is that a small excitation energy does not imply a small momentum transfer.
For nonrelativistic scattering, $Q_-\geq \omega_{q}/v$ as in Eq.~\eqref{eq:Qminbound}, so a GHz-scale transition can probe spatial scales much smaller than the overall qubit dimensions.
This makes the full finite-$Q$ response, rather than a conventional long-wavelength microwave matrix element, essential for quantitative particle-scattering predictions.

We also clarified how the target response scales in an array of qubits.
For independently prepared qubits in their product ground state, summing over the complete set of one-excitation final states gives the expected inclusive rate proportional to the number of qubits $N$, irrespective of whether the array is spatially coherent.
A genuine enhancement beyond this extensive scaling requires a correlated initial many-body state.
For symmetric Dicke states, the transition strength can scale as $(N-k)(k+1)$ and approach $N^2/4$ near the middle of the Dicke ladder, provided that the finite momentum transfer remains coherent over the participating array.
This establishes a clear distinction between simply increasing target exposure and engineering a quantum-enhanced scattering matrix element.

As a concrete application, we studied photon-mediated scattering of (nonthermally produced) bosonic superlight dark matter.
Using the phenomenological reference cross section $\bar{\sigma}_e$, we obtained projected sensitivities for realistic Xmon-inspired devices over the eV--keV mass range, which has been largely unexplored by the existing dark-matter direct-detection experiments, and quantified the impact of a representative residual dark-count background.
We also demonstrated that final-state Bose stimulation can significantly enhance the scattering rate for highly occupied bosonic dark-matter modes, leading to stronger projected sensitivity at the lowest masses than in the dilute-particle limit. This enhancement originates from the quantum statistics of the dark-matter ensemble and is distinct from the collective enhancement associated with entangled qubit states.
We further showed that frequency multiplexing provides a simple way to exploit the tunability of superconducting qubits:
for a fixed total exposure, distributing the array over several transition frequencies can extend the kinematic reach toward lower dark-matter masses while improving the sensitivity over a broad low-mass interval.

Taken together, these results identify superconducting qubits as more than passive microwave photon counters. 
They can act as quantum targets for direct particle scattering, with independently engineerable energy thresholds, spatial response, frequency coverage, and many-body quantum state.
The framework presented here provides the ingredients needed to combine microscopic particle models with realistic device electrodynamics and, ultimately, with quantum-enhanced detector architectures.

\begin{acknowledgments}
The work of DK is supported in part by the NSF Grant No. PHY-2609913.
The work of KK is supported in part by the  NSF Grant No. PHY-2609759. 
The work of MP is supported by the National Research Foundation of Korea (NRF) grant funded by the Korea government (MSIT) (No.RS-2025-00564488).
\end{acknowledgments}

\appendix

\section{Independent Qubits and Spatial Coherence}
\label{sec:independent_array}

Based on the formalism developed in the main text, one can define the transition current operator for a single qubit by
\begin{equation}
\hat{\mathcal{J}}^\mu(\vec q) = \int \dd^3x\,e^{i\vec q\cdot\vec x}\hat{J}_Q^\mu(\vec x)\simeq \int \dd^3x\,e^{i\vec q\cdot\vec x}J_Q^\mu(\vec x)(\hat a^\dagger -\hat a),
\end{equation}
where in the last step, the operator part is explicitly factored out.
Restricting the qubit Hilbert space to the \(|0\rangle\leftrightarrow|1\rangle\) transition, the raising part of the current operator is therefore
\begin{equation}
\hat{\mathcal J}_{10}^{\mu,+}(\vec q)
=
\mathcal{J}_{10}^\mu(\vec q)\hat\sigma^+,
\end{equation}
where the $\hat\sigma$ operator is assumed to govern the $\ket 0 \leftrightarrow \ket 1$ transition only. 
Considering \(N\) nominally identical qubits and assuming that the $A$-th qubit is centered at position \(\vec R_A\) (i.e., $\vec x$ is replaced with $\vec x+\vec R_A$), we find
\begin{equation}
\hat{\mathcal{J}}^{\mu,+}_N(\vec q)
=
\mathcal{J}^\mu_{10}(\vec q)
\sum_{A=1}^{N}
e^{i\vec q\cdot\vec R_A}\hat\sigma_A^+ .
\label{eq:array_current}
\end{equation}
The phase factor \(e^{i\vec q\cdot\vec R_A}\) specifies the relative phase with which a momentum transfer \(\vec q\) couples to different members of the array.

Suppose first that all qubits are initialized in the product ground state
\begin{equation}
|G\rangle = |0\rangle^{\otimes N} \equiv \ket{0_1 0_2 \cdots 0_N}.
\end{equation}
A scattering event can then excite any one of the \(N\) qubits.  Denoting by
\begin{equation}
|A\rangle
=
|0_1\cdots 1_A\cdots0_N\rangle 
\end{equation}
the state in which the \(A\)-th qubit is excited, the states \(|A\rangle\) are mutually orthogonal, i.e., \(\braket{B}{A}=\delta_{AB}\).  Consequently, after summing over the complete set of one-excitation final states,
\begin{equation}
\sum_F
\left|
\langle F|\hat{\mathcal J}^{\mu,+}_N(\vec q)|G\rangle
\right|^2
=
N\,|\mathcal J^\mu_{10}(\vec q)|^2 ,
\label{eq:independent_N}
\end{equation}
where \(|F\rangle\) runs over a complete orthonormal basis of the one-excitation final states, i.e., \(\ket F \in \{\ket A | A=1, \cdots , N \}\).
Thus, an array initialized in the product ground state has an inclusive microscopic scattering rate proportional to \(N\).

Importantly, this conclusion remains true even in the long-wavelength limit
\begin{equation}
Q|\vec R_A-\vec R_B|\ll1 ,
\label{eq:array_longwavelength}
\end{equation}
for which all phase factors in Eq.~\eqref{eq:array_current} become approximately equal.  The long wavelength makes the interaction spatially coherent across the array, but it does not make the distinct one-qubit final states identical.  The amplitudes therefore cannot simply be added and squared to obtain an \(N^2\) rate.

The same point can be seen by projecting onto the normalized symmetric one-excitation state,
\begin{equation}
|W_N\rangle
=
\frac{1}{\sqrt{N}}
\sum_{A=1}^{N}|A\rangle .
\end{equation}
In the limit of Eq.~\eqref{eq:array_longwavelength},
\begin{equation}
\langle W_N|
\sum_A\sigma_A^+
|G\rangle
=
\sqrt{N},
\end{equation}
so that even the bright collective channel carries a transition probability proportional to \(N\), rather than \(N^2\).  In this case, the oscillator strength of the \(N\) independent transitions is concentrated into a single symmetric final state, but the total strength is not increased.

This distinction is useful when comparing particle scattering with earlier proposals for quantum-enhanced searches for wave-like dark matter.  For example, Ref.~\cite{Chen:2023swh} showed that an appropriate quantum circuit can coherently combine the phase evolution induced on multiple sensor qubits by a classical oscillating dark-matter field, resulting in a signal probability proportional to \(N^2\) in the small-signal limit.  In that case, the enhancement arises from coherent phase accumulation and quantum-interference operations applied to the sensor qubits.  The situation considered in this work is different. An individual particle transfers a definite energy and momentum to the detector, and the relevant quantity is therefore the many-body transition matrix element of the operator in Eq.~\eqref{eq:array_current}.

We emphasize also that the exchanged photon in the process of interest is off shell and should not be regarded as an incident photon state whose coherence must be separately assumed.  For a specified momentum transfer \(\vec q\), the interaction itself carries the well-defined spatial phases appearing in Eq.~\eqref{eq:array_current}.  Whether different qubits participate coherently is instead controlled by the spatial resolution of the momentum transfer, Eq.~\eqref{eq:array_longwavelength}, together with the quantum state of the target.

\section{Collective Enhancement with Dicke States}
\label{sec:dicke_array}

A genuine enhancement beyond the extensive \(N\) scaling becomes possible if the array is prepared in a collective many-body state for which the transition matrix element itself grows with \(N\).  A convenient example is provided by the symmetric Dicke states~\cite{Dicke:1954zz}.  We denote the normalized state containing \(k\) excitations by
\begin{equation}
\left|D_N^{(k)}\right\rangle
=
\binom{N}{k}^{-1/2}
\sum_{\{A_1<\cdots<A_k\}}
|A_1,\ldots,A_k\rangle .
\label{eq:dicke_state}
\end{equation}
In the long-wavelength limit, Eq.~\eqref{eq:array_current} reduces to the usual collective raising operator,
\begin{equation}
\widehat \Sigma^+ = \sum_{A=1}^{N}\hat\sigma_A^+, \label{eq:collectiveRaising}
\end{equation}
which acts on a Dicke state according to
\begin{equation}
\widehat \Sigma^+
\left|D_N^{(k)}\right\rangle
=
\sqrt{(N-k)(k+1)}
\left|D_N^{(k+1)}\right\rangle .
\label{eq:dicke_matrix}
\end{equation}
The corresponding transition rate relative to a single-qubit ground-state transition is therefore
\begin{equation}
\frac{
\Gamma^{(N)}_{k\rightarrow k+1}
}{
\Gamma^{(1)}_{0\rightarrow1}
}
=
(N-k)(k+1).
\label{eq:dicke_rate}
\end{equation}

Equation~\eqref{eq:dicke_rate} makes clear why spatial coherence alone does not produce an \(N^2\) enhancement.  For the collective ground state, \(k=0\), one recovers
\begin{equation}
\Gamma^{(N)}_{0\rightarrow1}
=
N\,\Gamma^{(1)}_{0\rightarrow1},
\end{equation}
in agreement with Eq.~\eqref{eq:independent_N}.  By contrast, for a prepared Dicke state near the middle of the ladder, \(k\simeq N/2\),
\begin{equation}
(N-k)(k+1)
\simeq \frac{N^2}{4},
\end{equation}
up to subleading terms in \(N\)~\cite{Dong:2025mdk}.  The resulting enhancement is analogous to the collective matrix-element enhancement familiar from superabsorbing quantum systems~\cite{Higgins:2014cxq}.  In the present context, the purpose of the collective state is therefore not to establish the coherence of the exchanged interaction, but to engineer an initial many-body state whose transition matrix element is parametrically larger than that of a collection of independently prepared qubits.

For particle scattering, however, the finite momentum transfer cannot in general be neglected.  
We then need to consider the full form of the collective raising operator as in Eq.~\eqref{eq:array_current}: 
\begin{equation}
\widehat \Sigma^+(\vec q)
=
\sum_{A=1}^{N}
e^{i\vec q\cdot\vec R_A}\hat \sigma_A^+
\end{equation}
Starting from \(\left|D_N^{(k)}\right\rangle\) and summing over the complete set of \((k+1)\)-excitation final states gives
\begin{equation}
\frac{
\Gamma^{(N)}_{k\rightarrow k+1,{\rm incl}}(\vec q)
}{
\Gamma^{(1)}_{0\rightarrow1}(\vec q)
}
=
(N-k)
\left[
1+
\frac{k}{N-1}
\left(
N C_N(\vec q)-1
\right)
\right],
\label{eq:finiteq_dicke}
\end{equation}
where $C_N$ denotes the associated array coherence factor---which originates from the interference between the phase factors $e^{i\vec q \cdot \vec R_A}$---defined as
\begin{equation}
C_N(\vec q)
\equiv
\left|
\frac{1}{N}
\sum_{A=1}^{N}
e^{i\vec q\cdot\vec R_A}
\right|^2 .
\label{eq:array_coherence}
\end{equation}
In the long-wavelength or full coherence limit, \(C_N\rightarrow1\), hence Eq.~\eqref{eq:finiteq_dicke} reproduces the Dicke enhancement of Eq.~\eqref{eq:dicke_rate}.  When the relative phases are effectively randomized, \(\langle C_N\rangle\simeq1/N\), the interference terms vanish and the rate approaches the incoherent result \(N-k\).

The finite-\(Q\) condition is particularly important for the scattering process considered here.  Since the exchanged photon is off shell, its spatial momentum \(Q\) is not fixed by the transition energy \(\omega_{q}\).  Indeed, for nonrelativistic scattering the kinematics in Eq.~\eqref{eq:Qminbound} implies a minimum momentum transfer of order
\begin{equation}
Q_{\rm min}\sim \frac{\omega_{q}}{v},
\end{equation}
which can be much larger than \(\omega_{q}\).  A microwave-scale qubit transition therefore does not by itself guarantee coherence across a macroscopic qubit array.  The collective \(N^2\) limit requires both a suitably prepared many-body state and
\begin{equation}
Q D\lesssim1,
\label{eq:array_size_condition}
\end{equation}
where \(D\) characterizes the spatial extent of the coherently participating qubits.

The \(N^2\) regime should consequently be viewed as a possible quantum-enhanced operating mode. Realizing the collective enhancement would additionally require the preparation and preservation of the appropriate Dicke state, sufficiently uniform transition frequencies, and an architecture compact enough to satisfy Eq.~\eqref{eq:array_size_condition} over the momentum-transfer range relevant to the signal.

\section{Numerical Construction of the Xmon Electrostatic Response}
\label{app:xmon_response}

\begin{figure}[t]
    \centering
    \includegraphics[width=0.48\textwidth]{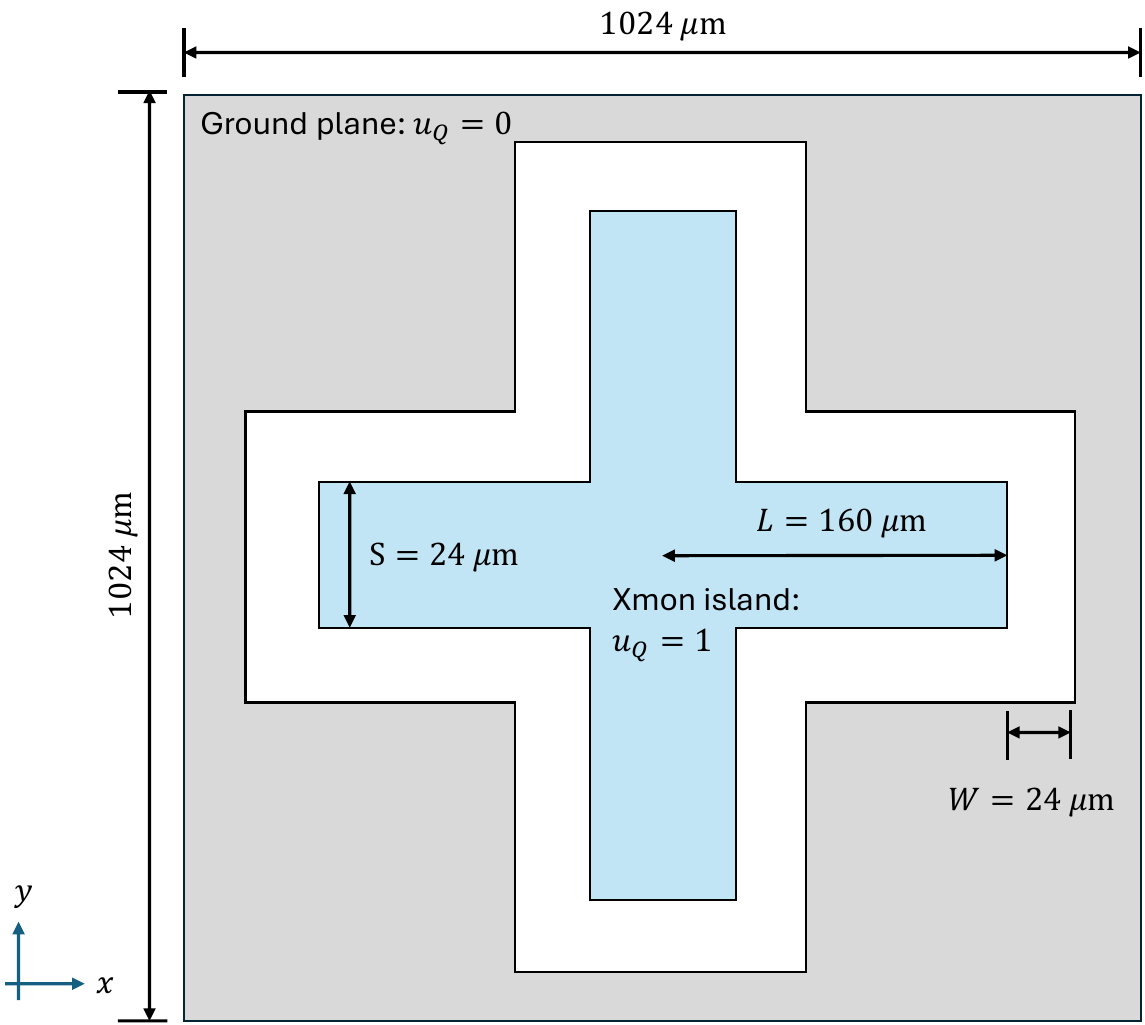}
    \caption{Schematic geometry used in the numerical construction of the Xmon electrostatic response. Dimensions are indicated for reference; essential components of the geometry are shown schematically and are not drawn to scale.}
    \label{fig:Xmonpicture}
\end{figure}

Here we provide the numerical prescription used to construct the
finite-momentum electrostatic response of the Xmon-style transmon
employed in Sec.~\ref{sec:applicationsI}.  
We model the Xmon as an infinitesimally thin conducting structure in
the $z=0$ plane (see Fig.~\ref{fig:Xmonpicture} for a schematic view). The island consists of the union of horizontal and
vertical rectangular strips,
\begin{equation}
 \mathcal I =
 \left\{
 |x|\leq L,\quad |y|\leq S/2
 \right\}
 \cup
 \left\{
 |y|\leq L,\quad |x|\leq S/2
 \right\},
 \label{eq:xmon_island_mask}
\end{equation}
where $L$ denotes the distance from the center of the cross to the end
of each arm and $S$ is the strip width.  The coplanar gap is modeled by
expanding each strip laterally and longitudinally by a distance $W$.
The region outside this expanded cross is taken to be a grounded
coplanar conductor.

For the benchmark used in this work, we employ one of the benchmark geometric parameters for the Xmon qubit capacitors in Ref.~\cite{Barends:2013zvp}:
\begin{equation}
 L=160~\mu{\rm m},
 \qquad
 S=24~\mu{\rm m},
 \qquad
 W=24~\mu{\rm m}.
 \label{eq:xmon_benchmark_geometry}
\end{equation}
The chip plane is represented by a $1024~\mu{\rm m}\times1024~\mu{\rm m}$ square with a uniform grid spacing
\begin{equation}
 \Delta x=\Delta y=1~\mu{\rm m}.
 \label{eq:xmon_grid}
\end{equation}
We assume that the Xmon island is held at unit potential while the surrounding coplanar ground plane is held at zero potential.

For numerical calculation purposes, the planar conductor is discretized into square pixels, according to the grid spacing above.  Denoting by $q_j$ the charge carried by the $j$th conducting pixel, the potential on a conducting pixel centered at $\vec r_i$ is approximated by
\begin{equation}
 V_i =
 \frac{1}{4\pi\epsilon_0\epsilon_{\rm eff}}
 \sum_j G_{ij}q_j ,
 \label{eq:bem_potential}
\end{equation}
where, for distinct pixels,
\begin{equation}
 G_{ij}=\frac{1}{|\vec r_i-\vec r_j|}.
 \label{eq:bem_kernel}
\end{equation}
For the self-interaction of a square pixel, the singular kernel is
replaced by its exact cell average,
\begin{equation}
 G_{ii}
 =
 \frac{4\ln(1+\sqrt{2})}{\Delta x}.
 \label{eq:bem_self}
\end{equation}

The benchmark is treated as a planar effective-interface problem with
\begin{equation}
 \epsilon_{\rm eff}\sim 5.2.
 \label{eq:xmon_epsilon_eff}
\end{equation}
This effective permittivity value should not be interpreted as the microscopic dielectric constant of a particular bulk substrate, but as an illustrative choice. Nevertheless, \(\epsilon_{\rm eff}\sim5\) is physically motivated by the planar air-sapphire geometry, while its precise value is unimportant within the simplified effective-interface treatment.

The linear system defined by Eq.~\eqref{eq:bem_potential} is solved
for the conductor-pixel charges subject to
\begin{equation}
 V_i =
 \begin{cases}
  1, & \vec r_i\in\mathcal I,\\
  0, & \vec r_i\in\mathcal G,
 \end{cases}
 \label{eq:bem_boundary}
\end{equation}
where $\mathcal G$ denotes the grounded coplanar conductor. In our implementation the dense Coulomb matrix is not constructed explicitly.  Instead, matrix-vector multiplication is evaluated by zero-padding the $N\times N$ charge grid to a $2N\times2N$ grid and performing the Coulomb convolution with a two-dimensional fast Fourier transformation. 

For a unit potential difference, the total charge on the island gives
the transmon capacitance,
\begin{equation}
 C_\Sigma = Q_{\rm island}
 =\sum_{j\in\mathcal I} q_j .
 \label{eq:xmon_capacitance_bem}
\end{equation}
For the benchmark geometry above we obtain
\begin{equation}
 C_\Sigma \simeq 70~{\rm fF},
 \qquad
 \frac{E_C}{2\pi}\simeq 275~{\rm MHz}.
 \label{eq:xmon_capacitance_validation}
\end{equation}

The conductor charge distribution obtained from the electrostatic
solution is normalized to the island charge,
\begin{equation}
 f_{\rm BEM}(\vec q_\parallel)
 =
 \frac{1}{Q_{\rm island}}
 \sum_j q_j
 e^{i\vec q_\parallel\cdot\vec r_j},
 \label{eq:xmon_charge_fourier}
\end{equation}
where $\vec q_\parallel=(q_x,q_y)$ is the component of the transferred momentum parallel to the chip plane.  Reversing the sign in the exponential gives the complex conjugate and therefore does not affect the squared response used below.
Here the function \(f_{\rm BEM}\) can be understood as the Fourier transform of the normalized conductor-charge distribution obtained from the electrostatic solution, playing a role analogous to the geometric charge form factor \(f_\rho\) in the idealized description of Sec.~\ref{sec:theoryresponse}.

Because each boundary-element-method (BEM) variable represents the total charge of a finite
square pixel rather than a point charge, we include the pixel form
factor
\begin{equation}
 P_{\rm pix}(q_x,q_y)
 =
 {\rm sinc}\left(\frac{q_x\Delta x}{2}\right)
 {\rm sinc}\left(\frac{q_y\Delta x}{2}\right),
 \label{eq:pixel_form_factor}
\end{equation}
where ${\rm sinc}(x)\equiv{\sin x}/{x}$.
In the effective homogeneous-dielectric representation used here,
the momentum-space device response corresponding to the electrostatic
construction of Sec.~\ref{sec:devicelevel} can be written as
\begin{equation}
 \widetilde\kappa_Q(\vec q)
 =
 -\frac{C_{\rm nat}}
 {\epsilon_{\rm eff} Q^2}\,
 f_{\rm BEM}(\vec q_\parallel)\,
 P_{\rm pix}(q_x,q_y),
 \label{eq:xmon_kappa_numerical}
\end{equation}
where
\begin{equation}
 C_{\rm nat}\equiv\frac{e^2}{2E_C},
 \qquad
 e^2=4\pi\alpha_{\rm em},
 \qquad
 E_C=\frac{e_{\rm SI}^2}{2C_\Sigma},
 \label{eq:xmon_cnat}
\end{equation}
where $e_{\rm SI}$ denotes the elementary electric charge, $e_{\rm SI}=1.602 \times 10^{-19}$~C, to distinguish it from the natural unit $e$.
For \(\Delta x=1~\mu{\rm m}\), the in-plane maximum resolved momentum (often called the Nyquist momentum) is
\begin{equation}
 q_{\rm Ny}
 =
 \frac{\pi}{\Delta x}
 \simeq 0.620~{\rm eV}.
 \label{eq:xmon_nyquist}
\end{equation}
We therefore perform the full numerical electrostatic calculation over
\begin{equation}
 0.003~{\rm eV}\leq Q\leq0.58~{\rm eV},
 \label{eq:xmon_qrange}
\end{equation}
where the upper limit is chosen to remain safely below the Nyquist momentum. 

Finally, the sensitivity calculation utilizes the device-orientation-averaged squared response in Eq.~\eqref{eq:dsigdq_device_avg}
\begin{equation}
 \left\langle
|\widetilde\kappa_Q(\vec q)|^2
\right\rangle_{\Omega_c}
 \equiv
 \frac{1}{4\pi}
 \int \dd\Omega_c\,
 \left|
 \widetilde\kappa_Q(\vec q)
 \right|^2 .
 \label{eq:xmon_orientation_average}
\end{equation}
Denoting $\theta_c$ and $\phi_c$ as the polar angle and azimuth measured relative to the
normal of the chip plane,
\begin{equation}
 q_\parallel=Q\sqrt{1-\cos^2\theta_c},~~~
 q_x=q_\parallel\cos\phi_c,~~~
 q_y=q_\parallel\sin\phi_c .
 \label{eq:xmon_qcomponents}
\end{equation}
We evaluate Eq.~\eqref{eq:xmon_orientation_average} using a midpoint
grid of
\begin{equation}
 N_{\theta_c}=24,
 \qquad
 N_{\phi_c}=48,
 \label{eq:xmon_angular_grid}
\end{equation}
with uniformly spaced points in
$\cos\theta_c\in[-1,1]$ and $\phi_c\in[0,2\pi)$.
Explicitly, the numerical quantity entering the rate calculation is
\begin{eqnarray}
  \left\langle
|\widetilde\kappa_Q(\vec q)|^2
\right\rangle_{\Omega_c}
 &=&
 \left[
 \frac{C_{\rm nat}}
 {\epsilon_{\rm eff}Q^2}
 \right]^2
 \frac{1}{N_{\theta_c} N_{\phi_c}} \\ \nonumber 
 &\times&  \sum_{a=1}^{N_{\theta_c}}
 \sum_{b=1}^{N_{\phi_c}}
 \left|
 f_{\rm BEM}(q_x^{ab},q_y^{ab})
 P_{\rm pix}(q_x^{ab},q_y^{ab})
 \right|^2 .
 \label{eq:xmon_response_discrete}
\end{eqnarray}

The full numerical calculation above is performed up to \(Q=0.58~{\rm eV}\). For the sensitivity calculation, which requires the response at larger momentum transfer for the highest dark-matter masses considered, we extend the device-orientation-averaged response beyond this range by extrapolating its high-\(Q\) behavior. A log-log fit to the directly calculated high-momentum tail gives
\begin{equation}
 \mathcal{K}(Q)=\left\langle
 |\widetilde\kappa_Q(\vec q)|^2
 \right\rangle_{\Omega_c}
 \propto Q^{-6.02},
\end{equation}
which we use to extend the response up to \(Q=5.8~{\rm eV}\). This procedure avoids the substantial computational cost associated with reducing the spatial grid size solely to resolve the high-\(Q\) tail. As a cross-check, we evaluated the response at several representative higher-momentum points using a finer spatial grid and found the extrapolated values to agree at the sub-percent level. Moreover, the extrapolated region gives only a small contribution to the total scattering rate, so the projected sensitivity is insensitive to the precise treatment of this high-\(Q\) tail.

\bibliographystyle{apsrev4-2}
\bibliography{refs.bib}

\end{document}